\documentclass{aa}
\usepackage{graphicx}
\usepackage{natbib}
\usepackage{wasysym}
\usepackage{txfonts}

\bibpunct{(}{)}{;}{a}{}{,}
    \def\void#1{{}}

    \def\be{\begin{enumerate}}
    \def\ee{\end{enumerate}}
    \def\bi{\begin{itemize}}
    \def\ei{\end{itemize}}

    \def\arcsec{$\, ^{\prime\prime}$}

    \def\~{$\sim$}
    \def\={$\simeq$}
    
    \def\H0{H_{0} \/$}
    \def\h0{H_{0} \/$}
    \def\h-1{$h^{-1} \/$}

\begin{document}
\title{The ATESP 5 GHz radio survey. \\
II. Physical properties of the faint radio population \thanks{Based on
observations carried out at the European Southern Observatory, La
Silla, Chile under program Nos.~75.A-0280 and 77.A-0211}}

\author{A.~Mignano\inst{1,2}, I.~Prandoni\inst{2}, L.~Gregorini\inst{3,2}, 
P.~Parma\inst{2}, H.~R.~de~Ruiter\inst{4,2}, M.~H.~Wieringa\inst{5}, 
G.~Vettolani\inst{6}, R.~D.~Ekers\inst{5}}

\offprints{A. Mignano, \email{amignano@ira.inaf.it}}
\institute{Dipartimento di Astronomia, Universit\`{a} di Bologna, 
via Ranzani~1, I-40126 Bologna, Italy
\and INAF - Istituto di Radioastronomia, Via Gobetti~101, I-40129 Bologna, 
Italy
\and Dipartimento di Fisica, Universit\`{a} di Bologna, via Irnerio~46, 
I-40126 Bologna, Italy
\and INAF - Osservatorio Astronomico di Bologna, via Ranzani~1, 
I-40126 Bologna, Italy
\and CSIRO Australia Telescope Facility, PO Box 76, Epping  NSW 2121, Australia
\and INAF, Viale del Parco Mellini 84, I-00136 Roma, Italy
}

\date{Received -; Accepted -}
\titlerunning{The ATESP 5 GHz radio survey. II. }
\authorrunning{A. Mignano et al.}

\abstract{

{\it Context: } One of the most debated issues about sub-mJy radio
sources, which are responsible for the steepening of the 1.4 GHz source 
counts, is
the origin of their radio emission. Particularly interesting, from this 
point of view, is the possibility of combining radio spectral index 
information with other observational properties to assess whether the sources 
are triggered by star formation or nuclear activity. 

{\it Aims:} The aim of this work is to study
the  optical and near infrared  properties of 
a complete sample of 131 radio sources with $S>0.4$ mJy, observed at both 1.4 
and 5 GHz  as part of the ATESP radio survey.
The availability of 
multi--wavelength radio and optical information is exploited to infer
the physical properties of the faint radio population.   

{\it Methods:} We use deep multi--colour (UBVRIJK) images, mostly taken in 
the framework of the ESO \emph{Deep Public Survey}, to optically
identify and derive
photometric redshifts for the ATESP radio sources.  
Deep optical coverage  and extensive colour information are available for 
3/4 of the region covered by the radio sample. 
Typical depths of the images are $U\sim 25$, $B\sim 26$, 
$V\sim 25.4$, $R\sim 25.5$, $I\sim 24.3$, 
$19.5\leq K_{s}\leq 20.2$, $J\leq 22.2$. We also add  
shallower optical imaging and spectroscopy obtained previously
in order to perform a global 
analysis of the radio sample. 

{\it Results:} Optical/near infrared counterparts are found for $\sim 78\%$ (66/85)
of the radio sources in the region covered by the deep multi--colour
imaging, and for 56 of these reliable estimates of the redshift and type  
are derived. We find that many of the sources with flat radio spectra 
are characterised by high radio--to--optical ratios ($R>1000$), typical of
classical powerful radio galaxies and quasars. 
Flat--spectrum sources with low $R$ values are preferentially identified with 
early type galaxies, where the radio emission is most probably triggered by 
low--luminosity active galactic nuclei. Considering both early type galaxies and quasars as 
sources with an active nucleus, such sources largely dominate our sample 
(78\%). Flat--spectrum sources associated with early type galaxies are quite 
compact ($d<10-30$ kpc), suggesting core-dominated radio emission.

\keywords{Surveys -- 
  Radio continuum: general - Methods: data analysis - Catalogues - Galaxies:
  general - Galaxies: evolution}} \maketitle

\section{Introduction}
\label{sec:introduction}

The faint (sub-mJy) radio population consists of a mixture of different
classes of objects. Since the early seventies it  has been known that the strongest 
sources are almost exclusively 
associated with either
active galactic nuclei (AGNs) or giant ellipticals, the latter of
which are also known as radio galaxies
(99\% above 60 mJy, \citealt{Windhorst90}). 
More recent work on mJy and sub-mJy sources has revealed that
faint sources are also found to be associated with normal elliptical, 
spiral and star-forming
galaxies, with the early type galaxies being the dominant component 
\citep{Gruppioni1999, 
Georgakakis1999, Magliocchetti2000, Prandoni2001b, Afonso2006}, while at 
$\mu$Jy levels
star-forming galaxies prevail (see e.g. \citealt{Richards1999}).

In spite of the progress made in our understanding of the faint radio 
population, many questions remain open.
For example, the relative fractions of the different types of objects 
are still quite uncertain, and our knowledge of their dependence on
limiting flux density is
still incomplete. The reason is, of course, that very little is known about 
the faint ends of
the various luminosity functions, and even less is known about the cosmological 
evolution of different kinds of
objects. This uncertainty is due to the incompleteness of optical 
identification and spectroscopy, since faint 
radio sources usually have very faint optical counterparts. 
Clearly {\it very}\, deep 
($I\apprge 25$) optical imaging and spectroscopy, for reasonably large deep 
radio samples, are critical if one wants to investigate these radio 
source populations.

Since the radio emission comes from different types of objects an
important question is what are the physical processes that trigger this emission. 
It is natural to assume that in the case of star-forming galaxies the
emission traces the history of galaxy formation and subsequent evolution by 
merging and interaction, while the emission in  AGNs will reflect black hole 
accretion history. To make matters more complicated, both processes may be 
present at the same time.

Although research in this field proceeds slowly due to very 
time--consuming spectroscopy much progress has been made in recent 
years thanks to strong improvement in the photometric 
redshift technique. 
Several multi--colour/multi--object spectroscopy surveys overlapping deep 
radio 
fields have recently been undertaken, including the Phoenix Deep Survey \citep{Hopkins1998,Georgakakis1999,Afonso2006} 
and the Australia Telescope ESO Slice Project (ATESP) survey 
\citep{Prandoni2000a,Prandoni2000b,Prandoni2001b}. In other cases,
deep multi--colour/multi--wavelength surveys have been complemented by deep 
radio 
observations (see e.g. the VLA--VIRMOS, \citealt{Bondi2003}; and the COSMOS, 
\citealt{Schinnerer2006}).

Multi--frequency radio observations are also important in 
measuring the radio spectral index, which may help to  
constrain the origin of the radio emission in the faint radio sources.
This approach is 
especially meaningful when high resolution radio images are available and 
radio source structures can be inferred. However, multi--frequency radio 
information is available for very few, and small, sub-mJy radio samples. 

The largest sample with multi--frequency radio coverage available so far 
is a complete sample of 131 radio sources with $S>0.4$ mJy, extracted from a 
square degree region observed at both 1.4 and 5 GHz as part of the 
ATESP radio survey 
\citep{Prandoni2000a,Prandoni2000b,Prandoni2006}. 

The $1.4-5$~GHz radio spectral index analysis of the ATESP radio sources 
was presented in the first paper of this series (\citealt{Prandoni2006}, 
hereafter Paper I). We found a flattening 
of the radio spectra with decreasing radio flux density. At mJy 
levels most sources have steep spectra ($\alpha \sim -0.7$, assuming 
$S\sim \nu^{\alpha}$), typical of synchrotron radiation, while at 
sub-mJy flux densities a composite population is present, with up to 
$\sim 60\%$ of the sources showing flat ($\alpha > -0.5$) spectra 
and a significant fraction ($\sim 30\%$) of inverted-spectrum ($\alpha>0$)
sources. This flattening at sub-mJy fluxes confirms previous results based on
smaller samples (\citealt{Donnelly1987,Gruppioni1997,Ciliegi2003}). 
Flat spectra in radio sources usually indicate the presence of
a self-absorbed nuclear core, but they can also be produced on larger scales 
by thermal emission from stars. 
 
It is possible to combine the spectral index information with other 
observational properties and infer the nature of the faint 
radio population. This is especially important with respect to the class 
of 
flat/inverted--spectrum sources as it permits us to study the 
physical processes that trigger  
the radio emission in those sources. This kind of analysis needs information 
about the redshifts and types of the galaxies hosting the radio sources. 
A detailed radio/optical study of the sample above is possible, thanks to the 
extensive optical/infrared coverage mostly obtained 
in the ESO \emph{Deep Public Survey} (DPS, 
\citealt{Mignano2007,Olsen2006}).

We give a brief discussion of all the data 
collected so far in Sect.~\ref{sec:datacoverage}, followed
by a more detailed analysis of the DPS optical data in 
Sect.~\ref{sec:dpsanalysis}, where we derive the UBVRI colour catalogue and 
photometric redshifts for the DPS galaxies in the region covered by the ATESP 
survey, assessing the reliability of the photometric redshifts themselves. In 
Sects.~\ref{sec:optid} and \ref{sec:radiozphot}, respectively, we
use the DPS UBVRIJK optical data to identify the ATESP 
radio sources and to derive photometric redshifts. 

A radio/optical analysis of 
the optically identified radio sources is presented in 
Sect.~\ref{sec:comp}, while in Sect.~\ref{sec:nature} we discuss the nature 
of the mJy and sub--mJy population on the basis of all the 
radio and optical data available to the ATESP sample. The main results are
briefly summarised in Sect.~\ref{sec:summary}.\\

Throughout this paper we use the $\Lambda$CDM model, with 
$H_0=70$, $\Omega_m=0.3$ and $\Omega_{\Lambda}=0.7$.

\section{Radio and optical data}
\label{sec:datacoverage}

\subsection{The ATESP 1.4 and 5 GHz surveys}
\label{sec:atesp}

As discussed  in Paper I, the Australia Telescope Compact Array (ATCA) was 
used to 
image, at 5~GHz, part of the $26^\circ\times 1^\circ$ strip of sky previously 
covered by the 1.4 GHz sub-mJy
ATESP survey (\citealt{Prandoni2000a,Prandoni2000b}). 
In the $2\times 0.5$ sq. deg. area 
observed at both 1.4 and 5 GHz a total of 131 distinct radio
sources are catalogued above a 6$\sigma$-threshold ($S>0.4-0.5$ mJy) 
at either 1.4 or 5 GHz (see Table 4 of Paper I). 
In particular we have 89 sources 
that appear in both the 1.4 and 5 GHz catalogues (\citealt{Prandoni2000b} 
and Paper I), while the remaining 42 
sources are catalogued only at one radio frequency: 20 sources at 1.4 GHz and 
22 sources at 5 GHz. For the sake of the spectral index analysis (see Paper I),
we searched for $3\sigma$ ($S\geq 0.2$ mJy) 
counterparts for these sources at the other radio frequency by directly inspecting the
(1.4 or 5 GHz) ATESP radio mosaics. 
As a result $\geq 3\sigma$ (1.4 or 5 GHz) flux 
measurements were provided for 29 additional sources (12 catalogued at 1.4 GHz
and 17 at 5 GHz), while for 13 (8 catalogued at 1.4 GHz and 5 at 5 GHz) 
sources (1.4 or 5GHz) $3\sigma$ upper limits were estimated.
Among the 131 sources catalogued at 1.4 and/or 5 GHz there are
three multiple sources: one is catalogued as triple at 1.4~GHz and as double 
at 5~GHz and another source is catalogued as double at 1.4~GHz and as a single 
non--Gaussian (extended) source at 5~GHz.

\subsection{The optical/infrared DPS survey}
\label{sec:dps}

The $2\times 0.5$ sq. deg. area imaged at both 1.4 and 5 GHz as part 
of the ATESP survey overlaps entirely with one sub-region (namely the DEEP1 
sub-region, see below) of the ESO DPS survey.
The DPS is a multi--colour survey consisting 
of both optical and near--infrared observations.
The DPS was carried out in the optical ($U,B,V,R,I$), using the WFI 
(Wide Field Imager) camera
mounted at the 2.2m ESO telescope, and in the NIR ($J,K_s$), 
using the SOFI camera mounted at the ESO NTT telescope. 
For a detailed description of the UBVRIJK filters used for the DPS 
we refer to  \citet{Mignano2007} and \citet{Olsen2006}.

The optical (UBVRI) observations cover three 
distinct $2\times 0.5$sq.~deg. regions of sky (named DEEP1, DEEP2, DEEP3). 
Each of the three regions is covered by four $0.5\deg\times 0.5\deg$ WFI 
pointings (a, b, c, d). Typical depths of the optical observations are 
$U_{AB}\sim 25.7$, $B_{AB}\sim 25.5$, $V_{AB}\sim 25.2$, $R_{AB}\sim 
24.8$, $I_{AB}\sim 24.1$ \citep{Mignano2007}. 

The infrared DPS comprises two strategies: shallow $K_s$--band 
($K_{s\,AB}\leq 21.3$) contiguous coverage of
about half the WFI fields, complemented by deeper $J$-- and $K_s$--band 
($J_{AB}\leq 23.4$ and $K_{s\,AB}\leq 22.7$) contiguous coverage 
($4\times 4$ SOFI pointings) of the central part of the WFI fields observed in 
the shallow strategy  \citep{Olsen2006}. In particular for region DEEP1, the
one of interest for this work, infrared coverage 
was proposed for  WFI fields DEEP1a and DEEP1b in shallow strategy 
($7\times 7$ SOFI pointings), and for the central part of them in deep strategy
($4\times 4$ SOFI pointings). 

There are some gaps in the optical/NIR imaging of  
the three fields of the region DEEP1 (see 
Table~\ref{tab:dps_completeness} for the summary of the observations).
Since NIR coverage of each single WFI field is obtained with several 
contiguous SOFI images, the seeing and the limiting 
magnitude values reported in the table for $J$ and $K_s$ bands 
are an average ($\pm$ standard deviation) of all the 
SOFI images contributing to the WFI field.
  
\begin{figure*}

\vspace{1.0cm}
\includegraphics[scale=0.3]{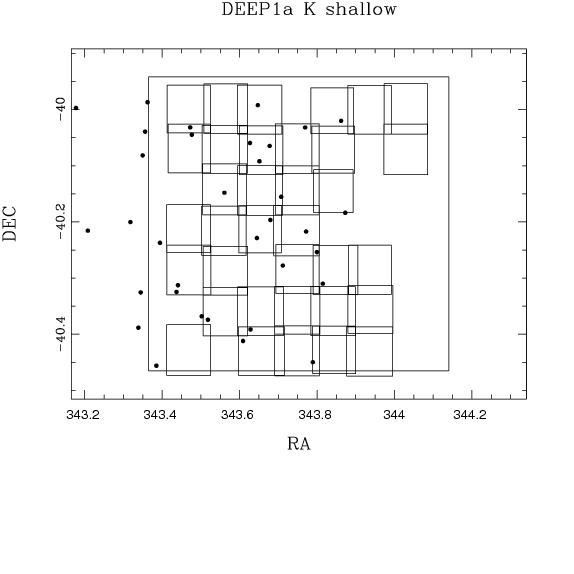}
\includegraphics[scale=0.3]{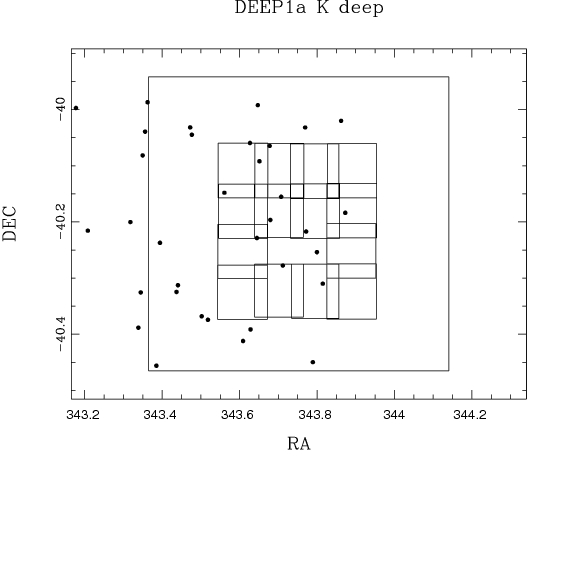}
\includegraphics[scale=0.3]{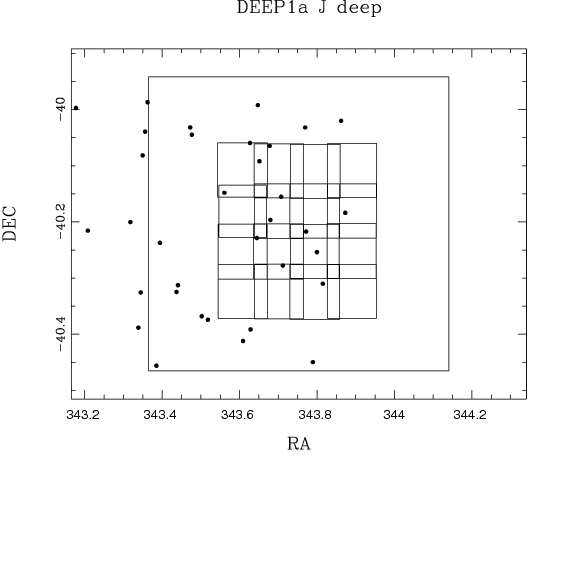}
\includegraphics[scale=0.3]{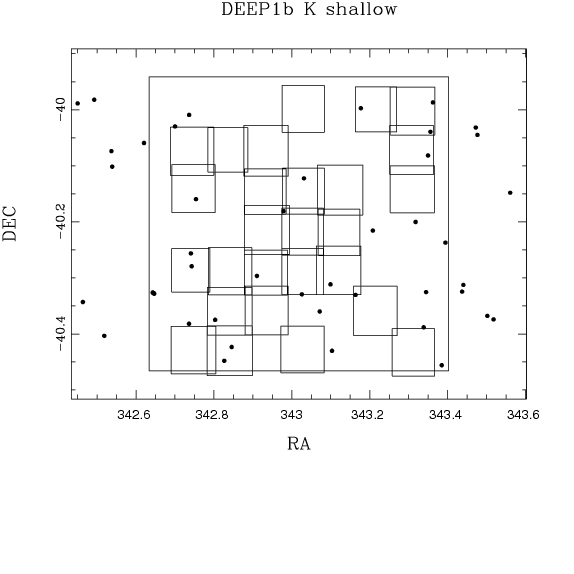}
\includegraphics[scale=0.3]{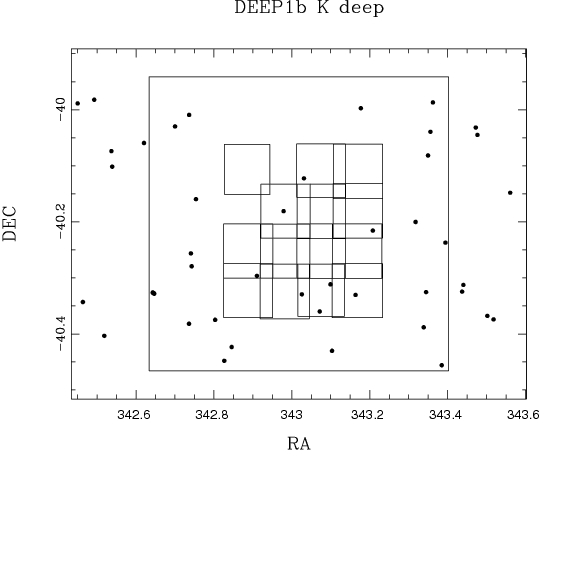}
\includegraphics[scale=0.3]{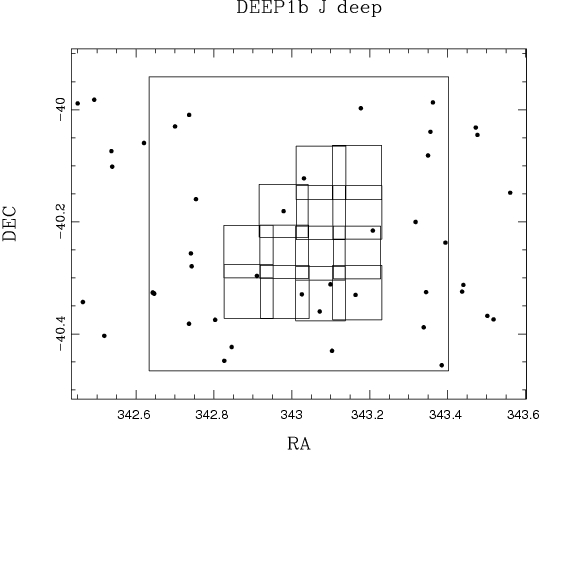}

\vspace{-1.0cm}
\caption{Optical/NIR data coverage for DEEP1a (top) and DEEP1b (bottom). From 
left to right coverage in the K-band (shallow and deep strategies) and J-band 
(deep strategy only).
\label{fig:completDPS}}
\end{figure*}

Figure \ref{fig:completDPS} shows the 
distribution of the infrared SOFI 
pointings over the WFI fields DEEP1a (top) and DEEP1b (bottom) 
for the two available infrared bands 
(J and K). $K-$band coverage is shown in both strategies: 
shallow and deep imaging (left and middle panels). $J-$band coverage 
(right panels) was obtained only for the deep strategy.
The infrared frames are represented by 
the small numbered squares that overlap the corresponding optical WFI frames 
(big squares). 

From Table~\ref{tab:dps_completeness} and Fig. \ref{fig:completDPS} 
it is clear that DEEP1a imaging is complete in the optical U, B, and R 
pass-bands, while no imaging is available in the V-band. 
The $K_s$--band imaging, on the other hand, is 70\% complete in the shallow 
strategy and 75\% complete in the deep strategy, while 100\% completeness is 
reached by the $J$--band imaging. 

The optical imaging of DEEP1b is complete, except for the $U$--band imaging, which
is slightly shallower than planned ($m_{lim} =\sim 24.6$). 
The deep infrared imaging has a good 
coverage in both filters ($>$80\%), while the shallow $K_s$--band imaging 
covers only about 55\% of the area.

It is interesting to note that, even if not complete, the infrared coverage of 
DEEP1a and b is distributed in such a way that many of the ATESP radio sources 
(filled black points in Fig.~\ref{fig:completDPS}) in the two fields 
(27 and 26 radio sources for DEEP1a and b respectively) have infrared information.
In particular, 75\% ($40/53$) of the sources have shallow $K_s$--band coverage, while 
deep $J$-- and $K_s$--band infrared data are available for 100\% ($20/20$) 
of the radio sources located in the central part of the fields.

DEEP1c was only observed in the V band (only down to $m_{lim}\sim 25$) and 
R band, while no observations are available for field DEEP1d.

Reduced images and single pass-band source catalogues extracted from both the 
optical and infrared DPS are described in detail in \cite{Mignano2007} and 
\cite{Olsen2006}, respectively, and are publicly available at the Centre de
Donn\'ees astronomiques de
Strasbourg (CDS).

\begin{table}[t]                      
\centering         
\caption{DPS optical and infrared data status and main attributes.
The table gives in Col. 1 the WFI field, in Col. 2 the pass-band, in Col. 3 
the seeing, and in Col. 4 the limiting 
magnitude ($5\sigma$, 2\arcsec aperture, Vega system). 
}  
\label{tab:dps_completeness}                                         
\begin{tabular}{lccc}                       
\hline                                     
\hline                                     
Field & Pass-band        & Seeing ($\prime\prime$)  & m$_{lim}$  \\     
\hline                                     
DEEP1a & $U$              &  1.37          &   25.26      \\       
       & $B$              &  1.37          &   25.85     \\      
       & $R$              &  0.87          &   25.74     \\        
       & $I$              &  0.86          &   23.76      \\     
       & $J$              & $0.676\pm 0.094$  & $22.17\pm0.23$ \\  
       & $K_{s,deep}$   & $0.712\pm0.090 $    &  $20.07\pm0.23$  \\   
       & $K_{s,shallow}$& $1.275\pm0.066 $    &  $19.57\pm0.16$   \\  
\hline                                                                      
DEEP1b & $U$              &  1.17          &   24.62   \\      
       & $B$              &  1.43          &   25.66   \\     
       & $V$              &  1.31          &   25.35   \\     
       & $R$              &  1.29          &   25.32   \\     
       & $I$              &  0.97          &   24.19   \\     
       & $J$              & $0.073\pm0.238$ &   $22.14\pm0.23$  \\   
       & $K_{s,deep}$   & $0.911\pm0.208$  &   $20.24\pm0.31$  \\  
       & $K_{s,shallow}$& $0.890\pm0.198$  &   $19.38\pm0.29$  \\  
\hline                                       
DEEP1c & $V$            &  1.19       &  25.03    \\       
       & $R$            &  0.98       &  25.43    \\       
\hline
\end{tabular}                                                
\end{table}                                                     

\begin{table}[t]                      
\centering          
\caption{Main attributes of additional optical imaging obtained for fields 
DEEP1a and DEEP1c. Columns as in table \ref{tab:dps_completeness}}  
\label{tab:addimaging}                                         
\begin{tabular}{lccccc}                       
\hline                                     
\hline                                     
Field & Pass-band        &  Seeing ($\prime\prime$)& m$_{lim}$  \\     
\hline                                     
DEEP1a & $V$             &  0.98          &   25.76      \\ 
\hline
DEEP1c & $U$            &  1.09       &  25.07    \\
       & $B$            &  1.27       &  26.56    \\  
       & $I$            &  1.21       &  24.83    \\ 
\hline                                                                   
\end{tabular}                                                
\end{table}                                                     
 
\subsection{Additional optical imaging}
\label{sec:new-wfi}

Since the DPS was not completed, we have undertaken new WFI 
optical observations in order to collect the missing data necessary to have 
full colour information for region DEEP1, and hence for our ATESP 
radio sources.
In this framework we have obtained V--band imaging for DEEP1a, and U-, B-, 
I-band imaging for DEEP1c. All these new observations were taken in 
collaboration with
the group that developed the Garching-Bonn Deep Survey (GaBoDS) data reduction pipeline 
(\citealt{Schirmer2003,Erben2005}) and therefore these new data were reduced 
through that pipeline. The main attributes for this additional imaging 
are shown in Table~\ref{tab:addimaging}. 
We refer to \cite{Hildebrandt2006} for a 
detailed description of the data (both reduced images and single pass-band 
source catalogues). 
Our multi--colour analysis of ATESP 
radio sources can rely on full UBVRI information for DEEP1a, b and c 
(plus infrared information for most of the sources in DEEP1a and b). 

\subsection{Other optical information}
\label{sec:otherdata}

It is worth mentioning that other optical imaging and/or 
spectroscopic data are available.
The 26 square degree area covered by the ATESP survey was chosen to overlap 
with the region where \cite{Vettolani97}
made the ESP (\emph{ESO Slice Project}) redshift survey. They performed a
photometric and spectroscopic study of all galaxies down to $b_J 
\sim$ 19.4. The ESP survey yielded 3342 redshifts (\citealt{Vettolani98}), 
to a typical depth of $z=0.1$ and a completeness level 
of 90\%. 

In the same region lies the \emph{ESO Imaging Survey} (EIS) Patch~A 
($\sim 3^{\circ}\times 1^{\circ}$ square degrees, 
centred at $22^h 40^m$, $-40^{\circ}$), mainly consisting of images in the 
I-band, out of which a galaxy catalogue ($95\%$ complete to $I=22.5$) 
was extracted (\citealt{Nonino99}). 
This catalogue allowed us to identify $\sim 57\%$ of the 386 
ATESP sources present in that region and optical spectroscopy was obtained 
for a complete magnitude-limited ($I<19$) sub-sample of 70 sources (see 
\citealt{Prandoni2001b}). Some VLT/NTT spectroscopy is also available for 
fainter sources in the same region ($\sim 40$ sources with $19<I<21.5$, 
Prandoni et al., in prep.). However, the 3 square degree ATESP-EIS 
sample only overlaps partially with the DEEP1 region, covering 
the fields DEEP1c and DEEP1d. 

This paper mainly focuses on the 
radio/optical analysis of the 85 ATESP radio sources located in DEEP1a, b 
and~c, for which  deep multi--colour optical/NIR information can be exploited. 
However, whenever considered useful, we include in our 
discussion any optical data (imaging and/or 
spectroscopy) available to the ATESP sources located in DEEP1d. Such data
may come either from the observations mentioned above, or from the literature.

\section{Multi--colour analysis of DEEP1 DPS data}
\label{sec:dpsanalysis}

A general discussion of the DPS optical imaging is provided
in \cite{Mignano2007} and in \citet{Hildebrandt2006}, where
the global quality of the photometry obtained through the EIS and the 
GaBoDS pipelines, respectively, is discussed. 
Here, we focus our attention on region DEEP1, which is the region of 
interest of this work. A careful analysis of the photometry of the 
single pass-band images covering region DEEP1 is very important since 
we will use this data later to estimate photometric redshifts 
for the ATESP radio sources. Also very important is the recipe followed to 
produce the optical colour catalogues, since reliable galaxy colours are 
crucial to get reliable photometric redshifts.  

\subsection{Colour catalogues}
\label{sec:colorcat}

We used the available UBVRI images to derive overall optical colour 
catalogues for DEEP1a, b and c. 

To obtain a good quality colour 
catalogue it is clear that one should use UBVRI images reduced in a consistent way. 
The optical images available to this work were reduced with 
different pipelines: the EIS pipeline for the images obtained in the 
framework of the DPS survey and the GaBoDS pipeline for the images obtained 
later on (UBI images for DEEP1c and V images for DEEP1a). 
In order to avoid internal inconsistency, we therefore decided to refer
to the EIS reduction for DEEP1b (see \citealt{Mignano2007}) and to the GaBoDS 
reduction for both DEEP1a and DEEP1c (see \citealt{Hildebrandt2006}). 

The technique of {\it reference imaging} (see below) was adopted in this work 
to produce the colour catalogue, 
since it provides the most accurate colour estimates, through the 
measuring of the source flux within the same area in any of the different 
pass-band images. This is especially important for extended objects.

We selected as the reference image the best seeing single pass-band image. Such a
choice allows us to minimise the effect of very close pairs of objects, which 
are not resolved due to poor seeing.
The $I$--band and R--band images for field DEEP1a
have very similar seeing values and the colour catalogue of DEEP1a was extracted
by using the $R$--band image as the reference. For DEEP1b and DEEP1c best
seeing is measured for I-- and R--band images, respectively, and the choice
of the reference image was done accordingly.

We run SExtractor (ver. 2.3, \citealt{Bertin1996}) in the 
so--called {\sc double image mode}:
detection and object apertures were based on the reference image,
followed by isophotal magnitudes  measured in the same aperture for each 
detected object on the other pass-band images separately.

The optical colour catalogues were then cross--correlated  with 
single pass-band catalogues extracted from
the $J$ and $K_s$ images, whenever available. Since the infrared  
catalogues overlap,
it may happen that the same optical object is identified
in more than one infrared catalogue. In such cases,
the infrared object with the lowest magnitude error was selected.  

We did not include the  NIR information in the colour catalogue
production from the beginning since a)
it is available only for limited sub-regions of DEEP1a and DEEP1b fields and 
b) the data are taken with a different instrument and telescope (SOFI at the 
3.6 m) and reduced through a specific EIS pipeline.

\subsection{Photometric redshifts}
\label{sec:photometric_z}

\begin{figure}[t]
\begin{center}
\includegraphics[scale=0.4]{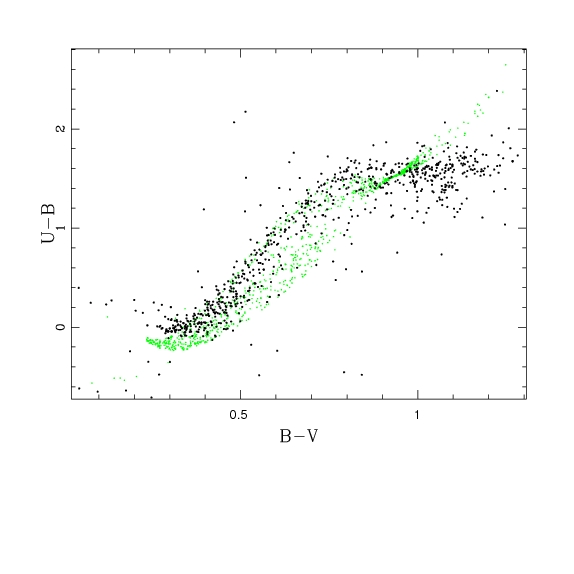}

\vspace{-1.5cm}
\includegraphics[scale=0.4]{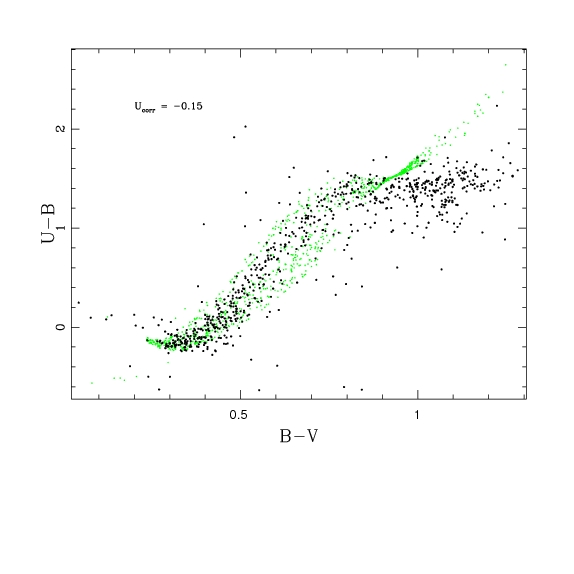}

\vspace{-1.5cm}
\caption{$U-B$ vs. $B-V$ colour diagram for stars in field DEEP1b. 
No correction applied (top), correction $U_{corr}=-0.15$ applied (bottom). 
Green points refer to DPS stars, black dots to modelled stars.}
\label{fig:color_checkUBBV}
\end{center}
\end{figure}

\begin{figure}[t]
\begin{center}
\includegraphics[scale=0.4]{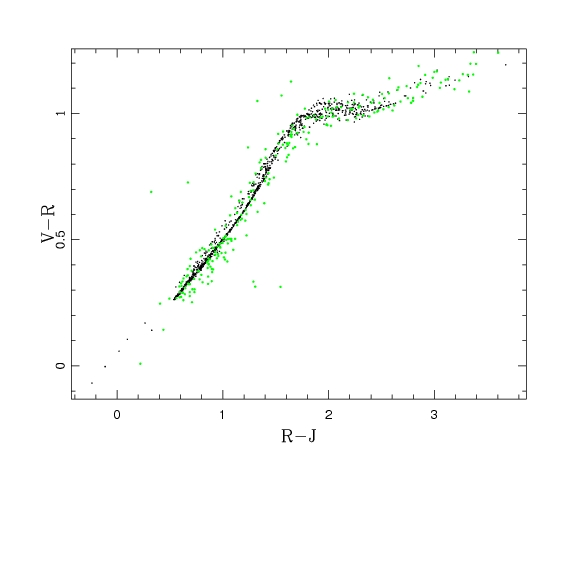}

\vspace{-1.5cm}
\includegraphics[scale=0.4]{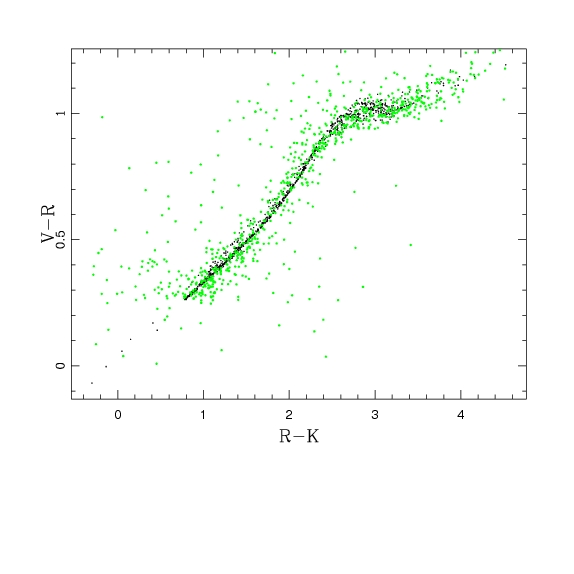}

\vspace{-1.5cm}
\caption{Optical vs. infrared colour-colour diagram for stars in DEEP1b: 
$V-R$ vs. $R-J$ (top) and $V-R$ vs. $R-K_s$ (bottom). 
Green points refer to DPS stars, black dots to modelled stars.}
\label{fig:color_checkVRRK}
\end{center}
\end{figure}

The success of photometric redshift estimate routines strongly depends
on the accuracy of the photometric calibration in the various pass-bands and 
on the accuracy of the colour estimation. In \citet{Mignano2007} and 
\citet{Hildebrandt2006}, comparisons between the UBVRI colours of stars in the 
various regions covered by the DPS, and the ones expected from a theoretical 
model \citep{Girardi2005}, were presented to check for the presence of possible
systematic offsets. Here, we report on the results obtained specifically for 
the DEEP1 region, which is the one of interest to our radio/optical study.
From the colour-colour diagram analysis very good agreement was found
between the catalogued star colours and the theoretical expectations, except in
the case of the  U-band for field DEEP1b, where an offset of $\sim 0.15$ mag is 
present (see Fig. \ref{fig:color_checkUBBV}, top panel). After correcting for
this offset, a good overlap between observed and expected colours is
obtained (see Fig. \ref{fig:color_checkUBBV}, bottom panel). 

We have also checked the optical (WFI)--infrared (SOFI) colours of the DPS 
stars, and no appreciable offset was seen.
This is shown in Fig.~\ref{fig:color_checkVRRK}, where $V-R$ vs. $R-J$ and $R-K_s$ 
are plotted for DEEP1b, chosen as reference.

We also used any spectroscopic data available from the literature
in this region to analyse the impact of both the correction 
applied in the U-band for DEEP1b and the use of NIR colours 
(when available) in the determination of galaxy photometric redshifts.  

It is important to note that most spectra come from the 
ESP redshift survey (see Sect. \ref{sec:otherdata}), which covers a limited redshift 
range ($z<0.3$, \citealt{Vettolani98}) and therefore the present comparison 
mainly probes the most local galaxies of the DPS. 

Photometric redshifts for 88 galaxies with spectroscopy information present 
in fields DEEP1a, b and c were estimated using the public
photometric redshift code {\it Hyperz} \citep{Bolzonella2000}, by using both 
the templates created from the synthetic stellar libraries of 
\citet{BruzualCharlot1993}, hereafter BC, and the empirical ones compiled by 
\cite{Coleman1980}  to represent the local galaxy population (hereafter CWW).
We stress that such galaxies are not necessarily associated to 
ATESP sources.

\begin{figure}[t]
\centering
\includegraphics[scale=0.4]{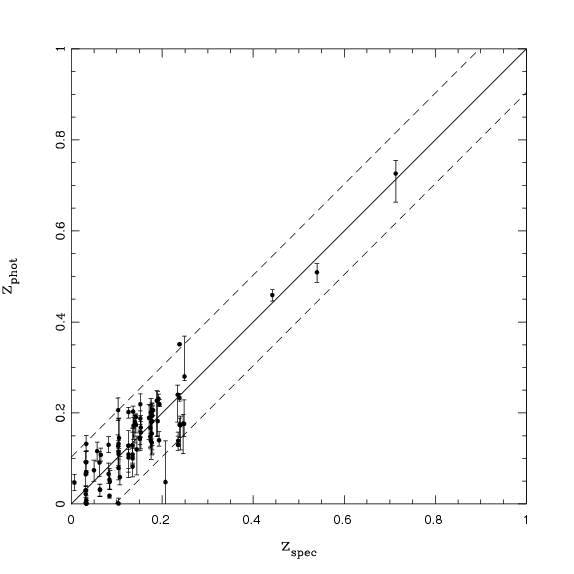}
\caption{z$_{phot}$ vs. z$_{spec}$ for the 88 galaxies in DEEP1a, b, c fields
with spectra available. 
$U$--band magnitudes of DEEP1b objects are corrected and 
NIR information is used, when available. The solid line indicates
$z_{phot}=z_{spec}$ and the dotted lines 
indicate the range that contains 95\% of the objects. 
The error bars represent the limits of the confidence intervals at 68\%}
\label{fig:ESPcheckcNIR}
\end{figure}

From this analysis we found a clear improvement in the photometric redshift
determination  when correcting for the 
systematic offset in U-band photometry. A further improvement is obtained 
when adding the NIR ($J$ and $K$) information (when available) 
to the $UBVRI$ colour catalogue. The $z_{phot}$ vs. $z_{spec}$ linear fit slope 
gets closer to unity ($a=0.93\pm 0.05$) and the object distribution around 
the $z_{phot} = z_{spec}$
line gets narrower. The final $z_{phot} - z_{spec}$ diagram is shown
in Fig.~\ref{fig:ESPcheckcNIR}. Dotted lines 
indicate the range that contains 95\% of the objects 
(z$_{phot}=$ z$_{spec} \pm$ 0.1). Such a range, albeit 
rather large, is adequate for this kind of study, where errors in luminosity
determinations of $\Delta logL$ of the order of $\la 0.5$ are acceptable. 
The horizontal error bars are not shown since they are 
negligible: ESP redshifts are characterised by errors of the order of 
$\sim 60$ km/s on the measured recession velocity, i.e. $\Delta z \sim 
2\cdot 10^{-4}$ (\citealt{Vettolani98}).

As a final remark, we stress that photometric redshifts shown in 
Fig.~\ref{fig:ESPcheckcNIR} 
were obtained using different template sets for different redshift ranges: 
galaxies with spectroscopic redshift $<0.1$ were fitted by CWW templates, 
while objects at z$_{spec}\geq 0.1$ by BC templates. This choice, as
expected, turned out to provide the best redshift estimates over the two 
redshift ranges.

\begin{figure}[t]
\begin{center}
\includegraphics[scale=0.43]{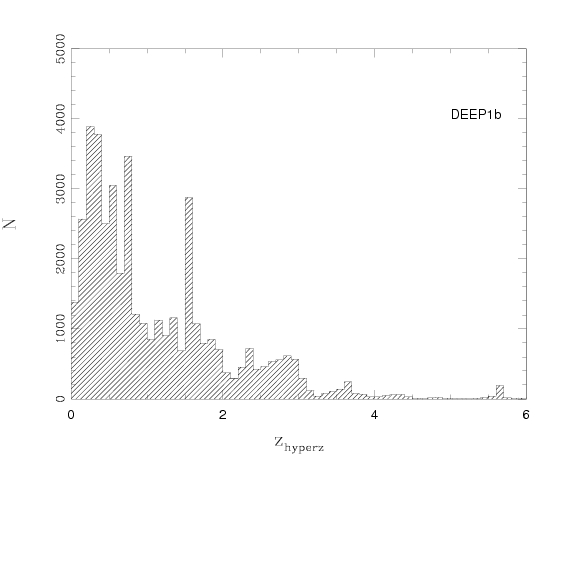}
\vspace{-1.5cm}
\caption{Galaxy photometric redshift distribution for field DEEP1b.}
\label{fig:zdistr}
\end{center}
\end{figure}

Figure~\ref{fig:zdistr} shows the galaxy photometric redshift distribution 
obtained from the optical UBVRI colour catalogues in field 
DEEP1b, chosen as reference.  
Most of the galaxies lie at $z<1$,
as expected, with a significant number of objects extending up to $z\sim 3$. 
On the other hand, the excess at $z\sim 5.5$ is mainly due to objects 
classified by {\it Hyperz} as Sc galaxies and is clearly spurious. For such 
objects the photometric redshift determination clearly fails.
Noteworthy are the two narrow peaks at $z\sim 0.7$ and $z\sim 1.5$. The 
latter is also present in fields DEEP1a and DEEP1c, and most probably 
indicates a degeneracy in the Hyperz routine, due to the fact that 
the spectral range covered by UBVRI-bands at $z\geq 1.5$ does not probe 
the Balmer $4000$ \AA\ break. More interesting is the peak at $z\sim 
0.7$, mainly composed of early type galaxies, which is not replicated in 
DEEP1a and c, possibly indicating the presence of real large scale 
structure. 

\section {Optical identification of the ATESP radio sources}
\label{sec:optid}

In the following we present the cross-correlation between the ATESP 
radio sources in fields DEEP1a, b and c and the multi--colour optical/NIR 
catalogues described in Sect.~\ref{sec:colorcat}.
 
In the literature different statistical techniques are used to cross--correlate
radio and optical catalogues, from the simplest, distance-only based 
criterion, which considers as \emph{good} any identification within a certain 
fixed radio--optical distance, to more sophisticated techniques, like 
the likelihood ratio criteria, based on the 
probability that a given source, at a certain 
distance and with a certain magnitude, is the true optical counterpart of the 
radio source (e.g. \citealt{deRuiter1977, Ciliegi2003, Sullivan2004, 
Simpson2006}).

In \cite{Mignano2007} a preliminary optical 
identification of the ATESP sources with the DPS catalogues was proposed, 
based on distance alone. However, while this choice proves to be
appropriate for shallower optical databases (see e.g. the ATESP-EIS case, 
\citealt{Prandoni2001b}), it is not very reliable when dealing, 
like in this case, with very deep 
(and therefore crowded) optical catalogues. 
Hence, we adopt the \emph{likelihood ratio} 
technique in the form described by \cite{Sutherland1992} and
\cite{Ciliegi2003}.

The likelihood ratio {\it LR} is defined as the ratio between the probability
that the source is the correct identification and the corresponding 
probability that the source is a background, unrelated object.
A threshold value $L_{th}$ of the likelihood ratio is assumed, above which 
a counterpart is 
considered as a good identification and below which is dismissed as spurious. 

The sample of accepted identifications thus consists of all the 
radio--optical associations that have $LR$ $>$ $L_{th}$. $L_{th}$ was 
chosen to be the value of $LR$ that maximises the 
function $(C+R)/2$, 
where $C$ is the completeness and $R$ the overall reliability
of the sample (\citealt{deRuiter1977}). 

\subsection {Optical identifications}
\label{subsec:idproc}

The ATESP radio sources were identified in the same reference pass-band 
(R or I) as chosen to derive the colour catalogues 
of DEEP1a, DEEP1b and DEEP1c. 

Before proceeding with the optical identification, the presence of possible 
systematic offsets between the radio and the optical astrometry was verified. 
We note that radio positions always refer to 5~GHz catalogue positions, 
unless the source is catalogued only at 1.4~GHz (i.e. 
$S_{\rm peak}(5\, $GHz)$<6\sigma$, see Paper I), while optical 
positions refer to the reference pass-band catalogue. 

As shown in \cite{Mignano2007}, where a preliminary analysis was
given, the median radio-optical offsets for our sample are 
$<\Delta$ RA$> = -0.$\arcsec213 and $<\Delta$ Dec$> = -0.$\arcsec073.
The source radio positions were corrected for such median offsets 
before proceeding with the radio source optical identification.

In computing the $LR$ value 
for each optical counterpart, 
the radio and optical positional 
uncertainties have to be taken into account. Here, we adopted $1 \sigma$ 
positional errors appropriate for the ATESP 
(see \citealt{Prandoni2000b}) and for the DPS (see \citealt{Mignano2007}) 
catalogues.
In addition, we need to assume an expected {\it a priori} identification rate 
($Q$). We adopted $Q= 0.7$, i.e. 
70\% of the radio sources are assumed to be truly identified down to the 
limiting magnitude of the optical catalogues. This choice is based on previous 
radio-optical identification studies undertaken down to similar optical 
depths (see e.g. \citealt{Ciliegi2005, Sullivan2004}). 

\begin{figure}[t]
\begin{center}
\includegraphics[scale=0.4]{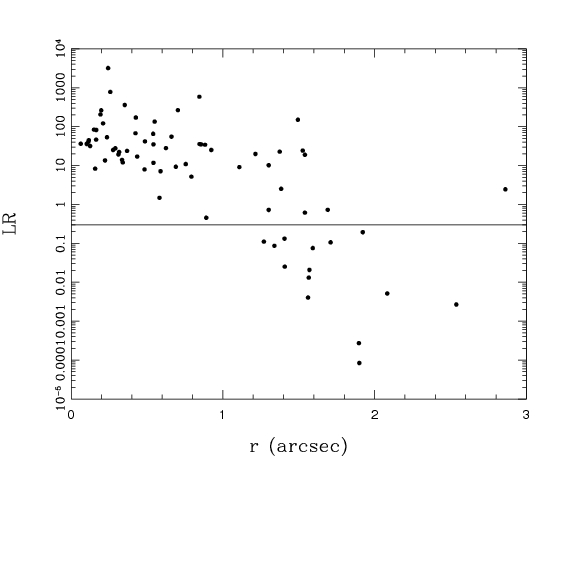}
\vspace{-1cm}
\caption{$LR$ values vs. radio--optical offsets for the 85 radio sources located
in DEEP1a, b and c. The horizontal solid line indicate the $LR$ value 
$L_{th}=0.3$, above which counterparts are considered as good identifications
(see text for details).}
 \label{fig:LRdistr}
\end{center}
\end{figure}

Figure \ref{fig:LRdistr} shows the distribution of $LR$ values as a function of 
radio--optical offsets for the 85 radio sources in DEEP1a, b, and c. 
As expected,
$LR$ decreases going to large radio--optical offsets ($>1$\arcsec), and the 
identifications become less reliable. The horizontal solid line represents 
the assumed threshold $LR$ value, above which optical 
counterparts are considered as good identifications. 
The adopted threshold value, $L_{th}=0.3$, 
was chosen in agreement 
with similar works reported in literature \citep[e.g.][]{Ciliegi2005}.
It is worth noting, however, that most of the sources have $LR$ values 
$\gg$ 10 (see Fig. \ref{fig:LRdistr}), which means that most of the optical 
identifications have very high probability of being real. 
As reported in Table~\ref{tab:LRtabstat}, 
60 radio sources in DEEP1a, b and c
were identified down to $L_{th}=0.3$, (see Col. 3).

In order to check the robustness of this identification technique and its 
dependence on the assumed parameters, the likelihood ratio analysis was 
repeated using different values of $Q$ in the range 0.5--1.0.
No substantial difference in the final number of identifications and in the 
associated reliability was found. 

The contamination due to possible spurious identifications with
$LR>L_{th}$ was estimated by shifting 
the coordinates of the radio sources by several random offsets and then 
repeating the identification procedure. 
The average contamination rate ($\%_{sp}$) was
7.4\%, 6.8\% and 6.3\% for 
DEEP1a, DEEP1b and DEEP1c, respectively (see Table~\ref{tab:LRtabstat}, Col. 5).

\tabcolsep 0.2cm
\begin{table}[t]
  \centering
  \caption{The sample identification statistics. Col.~1 lists the field name, Col.~2 the 
number of radio sources, Col.~3 the number of objects identified using the 
$LR$ technique, Col.~4 the completeness of the sample, Col.~5 the average 
contamination rate, Col.~6 the number of additional identifications, Col.~7 
the total number of identifications, and Col.~8 the identification rate.}
  \label{tab:LRtabstat}
  \begin{tabular}{lccccccc}
    \hline
    \hline
    Field & N$_{RS}$ & N$_{id}^{\geq L_{th}}$  &
    C & $\%_{sp}$ & N$_{id}^{add}$  & N$_{id}^{tot}$  & $\%_{id}$\\
\hline
DEEP1a & 27 & 16 &  98.6 & 7.4 & 4 & 20 & 74.1\\
DEEP1b & 26 & 21 &  99.1 & 6.8 & 1 & 22 & 84.6\\
DEEP1c & 32 & 23 &  99.0 & 6.3 & 1 & 24 & 75.0\\
\hline
 & \bf{85} & \bf{60}& \bf{98.9} & \bf{6.8} & \bf{6} & \bf{66} & \bf{77.6}\\
\hline
\end{tabular}
\end{table}

\subsection {Additional identifications}
\label{subsec-addid}

\begin{figure*}
\includegraphics[scale=0.32]{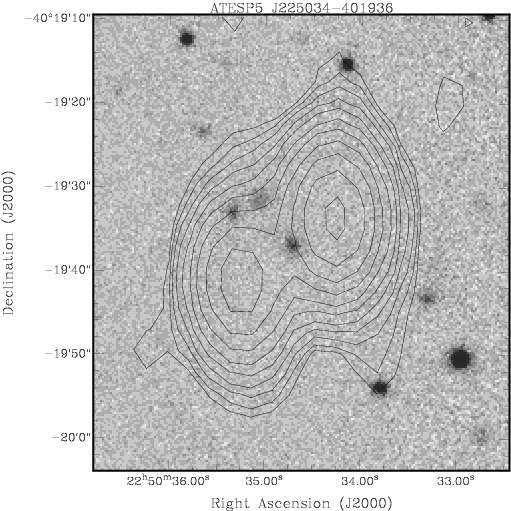}\includegraphics[scale=0.32]{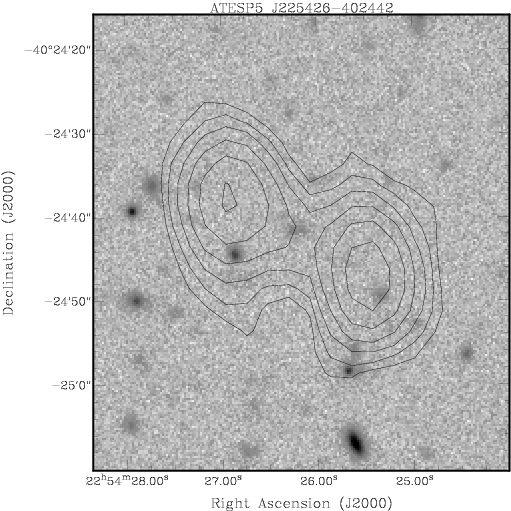}\includegraphics[scale=0.32]{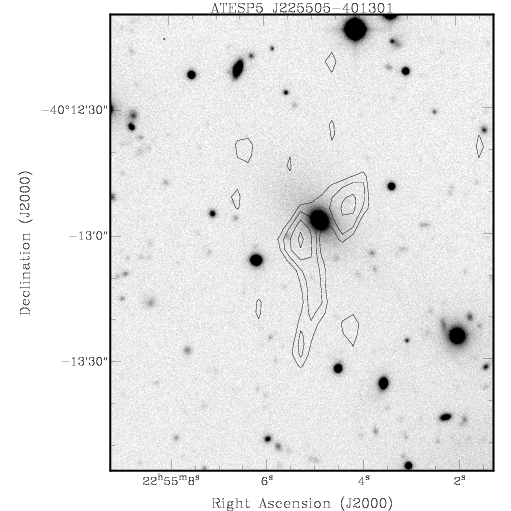}
\caption{The three extended radio sources included a posteriori in the 
identification sample: two with double radio morphology, 
ATESP5 J225034--401936 (left panel) and ATESP5 J225426--402442 (middle panel); 
and one wide angle tail source, ATESP5 J225505-401301 (right panel).
Grayscale: optical image in the reference band ($R$ for DEEP1a and $I$ for
DEEP1b). Contours: 5 GHz flux density.}
\label{fig:extended}
\end{figure*}

\begin{figure*}
\includegraphics[scale=0.32]{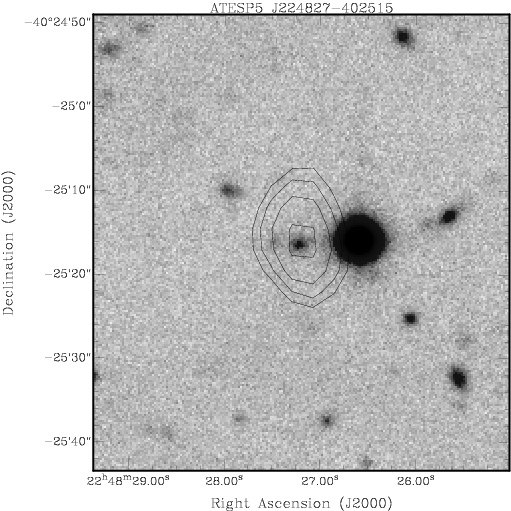}\includegraphics[scale=0.32]{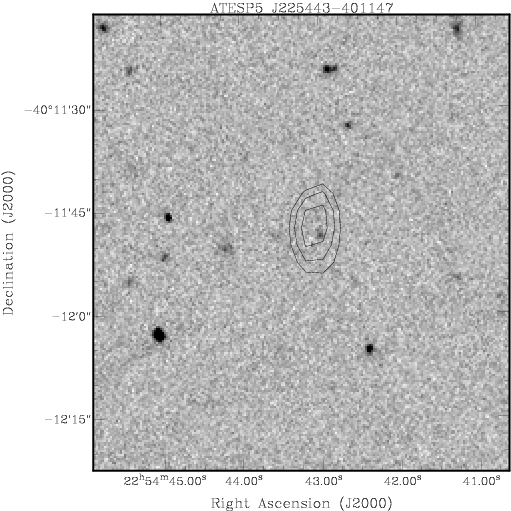}\includegraphics[scale=0.32]{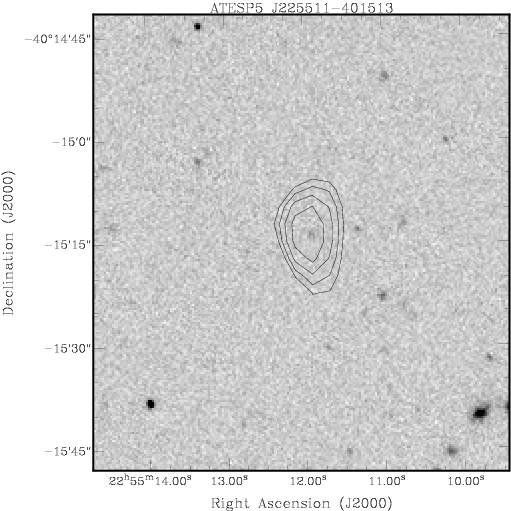}
\caption{The three sources identified in bands other than the reference one: 
ATESP5 J224827-402515 (left panel), identified in $I$--band;
ATESP5 J225443-401147 (middle panel), identified in $K_{s}$--band; 
and ATESP5 J225511-401513 (right panel), identified in $J$--band. 
Grey scale: optical or NIR image. Contours: 5 GHz flux density. }
\label{fig:nir}
\end{figure*}

The optical counterparts of the ATESP radio sources were all 
visually inspected on the corresponding optical reference-band images,  
giving particular attention to the 
multiple and non--Gaussian radio sources, where radio positions might not 
precisely coincide with the host galaxy core. From this inspection we 
decided to include among the identified radio sources three 
extended radio sources with  $LR$ values lower than the threshold: in all cases,
the optical counterpart located close to the radio barycenter is very likely to be
the host galaxy of the radio source. Two 
(ATESP5 J225034--401936 and ATESP5 J225426--402442) are classical
double radio sources (see Fig.~\ref{fig:extended}, left and middle panels) 
and one (ATESP5 J225505-401301) has the morphology of a low surface brightness 
wide angle tail (WAT) source, i.e. an extended radio source located in a 
cluster (see Fig.~\ref{fig:extended}, right panel). 
This hypothesis is supported by the 
fact that ATESP5 J225505-401301 is located in a crowded optical field 
with several optical galaxies in the field having similar photometric 
redshifts (see Sect.~\ref{sec:dpsanalysis} for 
the derivation of photometric redshifts for the optical sample).

In addition, we checked for any possible additional identification in
pass-bands other than the reference one ($R$ or $I$). We found that only
one extra identification (source ATESP5 J224827-402515 in DEEP1c field) 
could be recovered when the reference optical catalogue was extracted from the 
$I-$band image (see Fig.~\ref{fig:nir}, left panel). 
This identification was originally missed due to the fact that 
a larger region around the bright star close to the object was masked
in the $R-$band image.
This means that in general the reference images were chosen appropriately 
for our scientific application.

A similar check was performed for
the $J-$ and $K_{s}-$band infrared images, available for DEEP1a and b. 
For two sources (ATESP5 J225511-401513 and ATESP5 J225443-401147), with no 
optical counterpart in the DEEP1a optical images, a possible counterpart was 
found within a distance of 2\arcsec $\,$  in the infrared $K_{s}-$ or $J-$band 
images (see Fig.~\ref{fig:nir}, middle and right panels). 
These objects have extremely red colours ($R-K_{s}>5$), probably caused by 
either high redshifts or reddening due to dust. 

Including the six objects discussed above ($N_{id}^{add}$ in 
Table~\ref{tab:LRtabstat}), the final identification sample
is composed of 66 objects, corresponding to an identification rate of
77.6\% (see last column of Table~\ref{tab:LRtabstat}). 
On average the completeness $C$ is 98.9\% and the contamination rate is 6.8\%.
Both these quantities refer to the sample of 60
identifications, statistically defined on the base of the likelihood ratio 
technique. A summary of the sample identification statistics is given 
in Table~\ref{tab:LRtabstat}.

A list of all the identified radio sources is given 
in Table~\ref{tab:LRidentification}.
The six objects included a posteriori (see discussion 
above) are added at the bottom of the table.

A comparison with other similar radio/optical studies is shown in 
Table~\ref{tab:LRtabcomparison}. 
It is worth noting that the identification rate of our sample is 
consistent with the ones found in similar radio--optical studies taken from 
the literature. Of particular interest is the comparison with 
the identification rates reported for the 
VVDS--VLA sample \citep{Ciliegi2005}, and for the Phoenix survey 
\citep{Sullivan2004}, where the radio/optical analysis was performed down to
the same optical depth.

It is also interesting to compare the present study with the shallower
ATESP--EIS sample, where optical identifications were
searched down to $I=22.5$ (see \citealt{Prandoni2001b}).
The identification rate increases from $\sim57$\% of the ATESP--EIS to 
$78$\% of the ATESP--DEEP1, demonstrating the need for deep follow--up
surveys to properly identify the mJy/sub--mJy radio population.

\tabcolsep 0.2cm
\begin{table}[t]
  \centering
  \caption{Identification rate in our sample and in other deep radio fields: 
VVDS--VLA \citep{Ciliegi2005}, Phoenix survey \citep{Sullivan2004}, 
VLA--LH \citep{Ciliegi2003} and ATESP--EIS \citep{Prandoni2001b}. Col.~1 gives the 
sample name, Col.~2 the radio flux limit, 
Col.~3 the number of radio sources present in the sample, 
Col.~4 the area covered by the radio--optical data,
Col.~5 the limiting magnitude ($I$), and Col.~6 the 
identification  
rate. }
  \label{tab:LRtabcomparison}
  \begin{tabular}{lccccc}
  \hline
  \hline
  Survey   & S$_{lim}$ & N$_{RS}$ & Area          & I$_{lim}$ & \%$_{id}$ \\
           &  (mJy)    &             & (sq.degr.)   &          &      \\     
\hline
\bf{ATESP--DEEP1}&\bf{0.4}&\bf{85}   &  \bf{0.75}  & \bf{24.3}&   \bf{77.6}  \\
VVDS--VLA   & 0.08  & 1054       &  1         &      24.5     &   74.0    \\
Phoenix     &  0.1  & 839      &  3         &      24.5     &     79.0    \\
VLA--LH     & 0.05  & 63      &  0.03      &      24   &     92.0   \\
ATESP--EIS  & 0.4   & 386     &  3         &      22.5   &     57.3   \\
\hline
\end{tabular}
\end{table}

\section{Photometric redshifts for the identified ATESP radio sources}
\label{sec:radiozphot}

The optical spectra of radio sources are not exhaustively represented by the 
standard ``stellar'' templates used for normal inactive galaxies 
(Ellipticals, S0, Spirals, Irregulars, Star-bursts). 
We therefore added to the standard template spectra provided by 
{\it Hyperz} (BC and CWW) a set of spectral templates 
derived from the SDSS (Sloan Digital Sky Survey, \citealt{York2000}) quasar samples. 
In particular we added templates for: a) blue quasars (QSO, \citealt{Hatziminaoglou2000}); 
b) composite red quasars (REDQ, \citealt{Richards2003}); and c) composite broad 
absorption lines (BAL) quasars (BALQ, \citealt{Reichard2003b}). 
These templates are available at the SDSS web pages.

For each identified ATESP radio source {\it Hyperz} was used to provide a 
possible $z_{phot}$ (and corresponding reduced $\chi ^2$ probability) for each 
set of templates (CWW, BC, blue, red and BAL quasars). 
Then, the ``best'' (highest probability) z$_{phot}$ was selected as the correct
one, together with the corresponding spectral type. 

In eleven cases it was not
possible to assign a reliable z$_{phot}$ and spectral type to the optical 
radio source counterpart. 
Typically these are very faint objects
(mag$>$24 in detected bands), or objects with very limited colour information 
(e.g. detected only in NIR pass-bands), or objects with bad photometry due
to nearby star and/or deblending problems. One of these 
cases could be recovered thanks to the availability of spectroscopy 
information (source ATESP J224958-395855).

In summary, it was possible to assign a redshift and a spectral 
type to 56 of the 66 radio sources identified in DEEP1a, b and c (85\%).
However, if we restrict our analysis to a 
magnitude-limited $I<23.5$ complete sample, we get a success rate of 97\% 
(56/58 objects with redshift determination).

The relevant spectral parameters obtained for the 56 radio sources 
for which a redshift and type could be assigned, are reported in 
Table~\ref{tab:LRidentification}. 

Spectral types reported in Table \ref{tab:LRidentification} ( Col. 13) are defined as in 
\cite{Prandoni2001b}):

\begin{enumerate}
\item \emph{Early type spectra (ETS)}: ellipticals, early spirals (bulge--dominated 
Sa);
\item \emph{Late type spectra (LTS)}: late spirals (Sb, Sc, Sd) and irregular 
Magellanic (Im) galaxies; 
\item \emph{SB}: star-burst galaxy spectra (typical of HII regions);
\item \emph{PSB}: post star-burst galaxy spectra (K+A and E+A galaxies);
\item \emph{AGNs}: objects with evident characteristics of either 
Seyfert 1, Seyfert 2, or quasar spectra (respectively labeled as {\it Sy1, 
Sy2, Q});
\end{enumerate}

For the 14 objects with optical spectroscopy 
available, we have in general a good match between spectral and photometric 
redshifts ($\Delta z \la 0.1$, see Sect.~\ref{sec:photometric_z}).
The exception is ATESP J224803-400513, for which we have a $\Delta z >> 0.1$. In this
case, the obtained photometric redshift ($z_{phot}=1.0$) is much lower 
than the spectroscopic value of 1.72 (the published value, 
$z_{spec}=2.33$ by \citealt{Prandoni2001b}, was over-estimated). 

We also find very good agreement between photometric and spectral types. 
There is only one case (source ATESP J225400-402204) where the photometric type
(very old Sa) disagrees with the spectral type (LTS). 
However, note that passively evolving 
single bursts (defined as {\it Burst} in 
{\it Hyperz} BC templates) can be considered early or late, depending on 
their age. As a general rule, very old galaxies (age  
$\apprge 1$ Gyr) are included 
among the ETS, while very young galaxies (age $\apprle 0.1$ Gyr) are included 
among the LTS. For intermediate cases (ages between 0.1 and 1 Gyr) the 
classification is not straightforward, from wide-band 
information on the continuum shape alone. It is difficult to distinguish 
between LTS, ETS and PSB, with no information on
the presence of narrow absorption 
and/or emission lines. For the sake of simplicity we decided to make a sharp
separation between LTS and ETS at age $=0.3$ Gyr, with the caveat that 
among such objects we could have some mis--classification. The value of 0.3 
Gyr was chosen from a comparison between spectral type and 
{\it Burst} age in the few cases where spectroscopy was available.
One probably mis--classified object is source ATESP5 J225321-402317 
(Burst age 0.18 Gyr), which has linear 
size $\sim 200$ kpc and extended radio morphology, clearly indicating an 
AGN origin of its radio emission (see Fig.~\ref{fig:si_plots}, middle panel).

The final classification given to the optical counterpart on the base of
the present discussion is reported in last column of 
Table~\ref{tab:LRidentification}. The classes are defined following the 
spectral type definitions listed above. In the one case where photometric 
and spectral types disagree, we rely on the latter to define the object class.

To further check the reliability of our photometric redshift determinations,
we compared our redshift distribution with the one expected for ETS 
on the base 
of the well-known K--z correlation found for radio source host galaxies 
(see e.g. \citealt{Willott2003}) and with the R--z relation found for host 
galaxies of gigahertz peaked 
spectrum (GPS) radio sources (see \citealt{Snellen1996};
\citealt{Rigby2007}). We find that the photometric redshift distribution 
obtained for objects classified as ETS in our sample  
follows, within a $\Delta z\sim 0.1-0.2$ dispersion, the one expected on the 
base of the quoted relations. 

\section{The ATESP--DEEP1 source properties}
\label{sec:comp}

\begin{table}[t]
\caption{The ATESP sample composition.}
\label{tab:ATDEEP1comp}
\begin{tabular}{l|ccccc}
\hline
\hline
Sample & $I_{lim}$ & ETS & LTS+SB & AGNs & UNCL \\
 &  & (\%) & (\%) & (\%) & (\%)\\
\hline
ATESP--EIS & $19$ &   $49\pm8$  & $43\pm8$ & $9\pm3$  & $-$\\
ATESP--DEEP1 & $23.5$& $64\pm10$ & $19\pm6$ & $14\pm5$ & $3\pm2$\\
\hline
\end{tabular}
\end{table}

We exploited the photometric redshift and spectral type determinations for the 
ATESP sources in the DEEP1a, b and c regions,  to 
study the composition of the ATESP sample and the
radio/optical properties of mJy and sub--mJy sources. 
It is important to note that the ATESP--DEEP1 sample overlaps with the 
ATESP--EIS sample. This means that 
photometric (or spectroscopic) redshifts obtained for sources in DEEP1a, b 
and 
c could be in principle complemented by the sparse spectroscopic information 
available from the ATESP--EIS for DEEP1d sources. Nevertheless, in
the following it was preferred to limit our analysis to the sources
in DEEP1a, b, and c, 
(see Table~\ref{tab:LRidentification}), which represent a much more reliable 
sample, thanks to the very high identification/redshift determination 
statistics. 

Of the 66 identified radio sources, it was found that 37
are ETS, 8 are quasars  (AGNs), 10 are LTS, 1 is a SB, while  
10 objects could not be classified (UNCL). 

In Table \ref{tab:ATDEEP1comp}, the ATESP--DEEP1 composition is compared with 
the one found for the ``brighter'' ATESP--EIS sample 
(70 objects with complete 
spectroscopy down to $I=19$, \citealt{Prandoni2001b}). 
The ATESP--DEEP1 sample 
provides insight into the composition
of the faint radio population associated with optically faint galaxies, 
albeit in this comparison we restrict
our analysis to the magnitude-limited sample of 58 objects with $I<23.5$, 
to reduce the number of unclassified objects.
Table \ref{tab:ATDEEP1comp} shows that, as 
suggested by previous studies (e.g. \citealt{Gruppioni1999,Prandoni2001b}), 
the contribution of star--forming 
(LTS plus SB) galaxies decreases dramatically with the magnitude, going 
from 43\% of the 
ATESP--EIS ``bright'' ($I<19$) sample to 19\% of the ``deeper'' ($I<23.5$)
ATESP--DEEP1 sample. The fraction of ETS and AGNs, 
on the other hand, increases going to deeper magnitudes, even though 
the statistical uncertainties are large. 

\subsection{Redshift distribution}

\begin{figure}[t]
\centering
\includegraphics[scale=0.4]{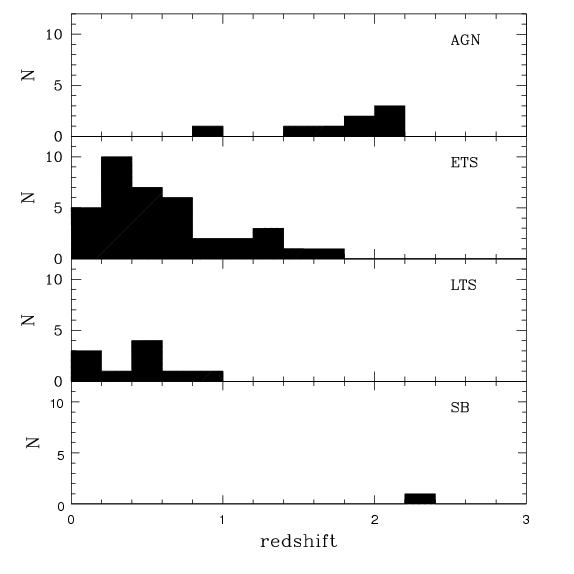}
\caption{Redshift distribution for the 56 radio sources in the ATESP--DEEP1 
sample with photometric redshift determination.
The sample is divided into four different classes. From top to bottom:
AGNs, ETS, LTS, and star-burst galaxies.}
\label{fig:radiozphotdistr}
\end{figure}

Figure \ref{fig:radiozphotdistr} shows the redshift
distribution of the  56 ATESP radio sources in regions DEEP1a, b and c, for 
which a reliable redshift estimate was obtained 
(see Table~\ref{tab:LRidentification}). Whenever spectroscopy
is available (14 objects), we rely on the spectral redshift determination.
The distribution of ETS shows a significant peak 
at z=0.4, with a tail extending up to z$\sim2 $, while, as expected, quasars 
have typically higher redshifts $1<z<2$, and LTS are 
found at z$<<$1.  
This reflects the fact that radio sources 
triggered by star formation are usually characterised by lower radio 
powers than sources triggered by AGN activity 
(see also Fig.~\ref{fig:radiopowerdistr}). 
The only star-burst galaxy in the sample has either a 
redshift ($z\sim 2$) or a radio power ($P_{\rm 1.4~GHz}$ close to 
$10^{26}$ W/Hz, see Fig.\ref{fig:radiopowerdistr}) that is much higher than 
for the 
LTS galaxy population. While this could be due
to evolutionary effects in the population of the radio--selected star forming
galaxies, it is also possible that the photometric classification is wrong.
In fact the SB spectra are notoriously similar to narrow--line  
AGN spectra (Seyfert 2) and photometric techniques based on wide--band colours 
could easily fail in classifying such objects. In addition, the photometric
routine applied to this sample ({\it Hyperz}) does not provide template 
spectra for Seyfert 2 galaxies. 
 
\subsection{Radio and optical luminosities}

\begin{figure}[t]
\centering
\includegraphics[scale=0.4]{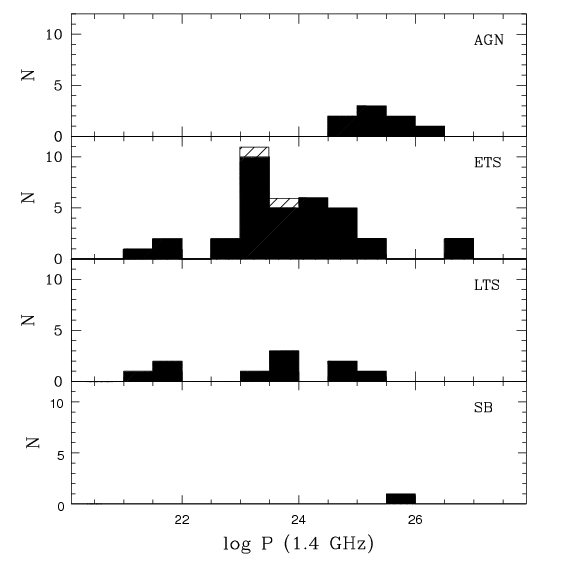}
\caption{1.4~GHz radio power distribution for the 56 ATESP--DEEP1 radio sources
with photometric redshift determination. 
The sample is divided into four different classes. From top to bottom: AGNs, 
ETS, LTS and star-burst galaxies. Light shading 
indicates two upper limits. }
\label{fig:radiopowerdistr}
\end{figure}

\begin{figure}[t]
\centering
\includegraphics[scale=0.4]{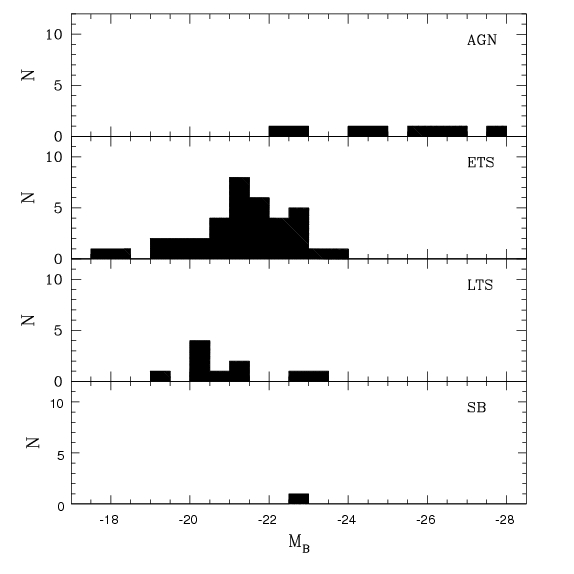}
\caption{Absolute $B$--band magnitude distribution. The sample is divided
into four different classes. From top to bottom: AGNs, ETS, LTS  and star-burst galaxies.}
\label{fig:Babsmagdistr}
\end{figure}

\begin{figure}[t]
\includegraphics[scale=0.4]{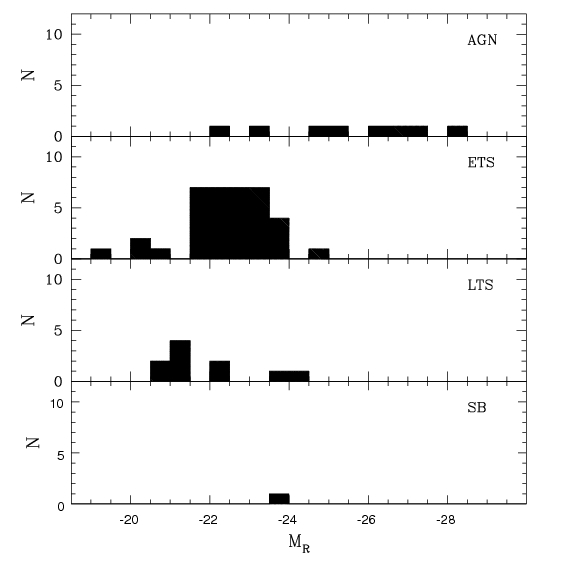}
\caption{Absolute $R$--band magnitude distribution. The sample is divided
into four different classes. From top to bottom: AGNs, ETS, LTS 
and star-burst galaxies.}
\label{fig:Rabsmagdistr}
\end{figure}

For the 56 ATESP radio sources in DEEP1a, b and c with redshift determination
we derived radio and optical/NIR  luminosities.
Radio powers were K--corrected by using the 1.4 -- 5~GHz radio spectral index 
of each source (see Table~\ref{tab:LRidentification}), 
while absolute magnitudes (computed by Hyperz) were K--corrected on the base 
of the optical spectral type. 

Figure \ref{fig:radiopowerdistr} shows the 1.4 GHz radio power distribution 
for the sample. Again, the four classes (AGNs, ETS, LTS, and SB) are shown 
separately.
ETS galaxies mostly have $23< \log P \rm \; (W/Hz) <25$, which are typical 
values of FRI 
radio sources \citep{Fanaroff1974}, while AGNs are, as expected,
characterised by higher radio powers ($10^{25}-10^{26}$ W/Hz).
LTS galaxies, on the other hand, have low radio powers,
with 7/10 having $P<10^{24}$ W/Hz, typical of radio sources triggered by star 
formation (see e.g. \citealt{Condon1988}).

If we assume that the ETS are triggered by low to medium 
luminosity AGN activity and put both AGNs and ETS objects in a single 
class, one finds that the sample is largely dominated by galaxies 
with an active nucleus (78\%, see Table~\ref{tab:ATDEEP1comp}), which
further demonstrates that sub--mJy samples like the ATESP are best suited to 
study the evolutionary behaviour of low--power AGNs. 

Figures~\ref{fig:Babsmagdistr} and \ref{fig:Rabsmagdistr} show the absolute 
magnitude distributions in B-- and R--bands for the 56 
ATESP--DEEP1 radio sources in fields DEEP1a, b and c with a redshift 
determination. AGNs are 
characterised by higher optical luminosities than ETS, LTS and SB galaxies. 
This is not surprising when we consider that in our sample all AGNs are 
photometrically and/or spectroscopically classified as quasars 
(see Table~\ref{tab:LRidentification}). 

\section{Nature of the mJy and sub-mJy radio population}
\label{sec:nature}

\begin{figure*}[t]
\resizebox{12cm}{!}{\includegraphics[]{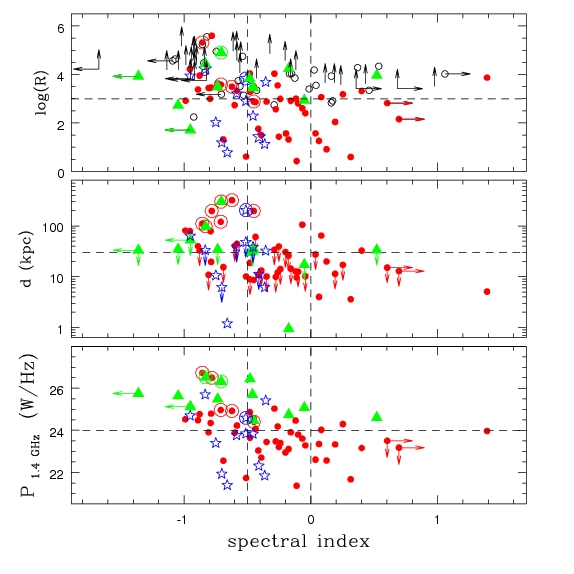}}
\hfill
\parbox[b]{55mm}{
\caption{Radio--to--optical 
ratio ($R$, top panel), linear radio
source size ($d$ in kpc, middle panel) and 1.4 GHz radio power
(in W/Hz, bottom panel) against 1.4 -- 5 GHz spectral index 
for the radio sources from the ATESP--DEEP1 sample: 
red filled circles (ETS); blue stars 
(LTS/SB); green filled triangles (AGNs). Circled
symbols indicate sources with extended and/or two-component radio morphology,
typical of classical radio galaxies. 
Black open circles are for identified sources that do not have a 
redshift/type determination. Arrows indicate upper/lower limits. 
Vertical dashed lines indicate the $\alpha =-0.5$ and the $\alpha =0$ 
values, above which source spectra are defined respectively 
as flat and inverted. Horizontal dashed lines in the three panels indicate, 
from top to bottom,
values of $R=1000$, $d=30$ kpc and $P_{\rm 1.4 GHz}=10^{24}$ W/Hz, respectively.}
\label{fig:si_plots}}
\end{figure*}

In order to probe the origin (nuclear or on a larger scale) of the radio 
emission in mJy and sub--mJy sources and the physical processes responsible 
for the flattening of the radio spectral index found in sub-mJy samples like
the ATESP (see \S\ref{sec:introduction}), we made an overall 
comparison of the radio spectral index, the radio morphology and the optical 
properties of the entire ATESP--DEEP1 sample.  

In Fig.~\ref{fig:si_plots} (top panel) the radio--to--optical ratio 
is plotted as a function of spectral index
for the whole ATESP--DEEP1 sample (fields a, b, c and d). 
The radio--to--optical ratio
was defined following \cite{Condon1980}, as $R=S\cdot 10^{0.4(m-12.5)}$, 
where $S$ is
the source 1.4 GHz flux density (in mJy) and $m$ is the optical magnitude 
(here 
we assume the I--band magnitude). 
We thus can include sources without known 
redshifts. In the following we use both DPS and 
ATESP-EIS optical data (see 
\citealt{Prandoni2001b}), when available, while lower limits to $R$ 
are given whenever a source was not 
identified down to the limiting magnitude of the optical surveys ($I\sim 22.5$ 
for EIS--WIDE and $I\sim 24$ for DPS DEEP1). 
For the sources with spectral type/redshift estimates available (either from
multi--colour photometry or spectroscopy) we can 
distinguish between ETS (red filled circles), LTS 
plus star-burst galaxies (blue stars) and AGNs (green double triangles). 

Figure~\ref{fig:si_plots} 
clearly shows that most of the flat--spectrum sources have 
high radio--to--optical ratios ($R>1000$), typically associated with the 
classical powerful radio galaxies and quasars.
Flat--spectrum sources with low $R$ values 
are preferentially identified with ETS, where the
radio emission is again probably triggered by nuclear activity 
(typical radio powers $P \sim 10^{23-25}$ W/Hz, 
see Fig.~\ref{fig:radiopowerdistr} and discussion therein). 
Star--forming galaxies (LTS and SB), on the other hand, are typically 
associated to steep--spectrum 
sources, as expected for synchrotron emission in galactic disks or in 
nuclear star-bursts. 

A further radio/optical analysis of the ETS in the ATESP--DEEP1
sample has shown that 
ETS with flat and/or inverted spectrum are preferentially 
compact (linear sizes $d< 10-30$ kpc, see Fig.~\ref{fig:si_plots}, middle
panel). Their rather low radio luminosities 
($P_{1.4 \rm{GHz}}\sim 10^{22-24}$ WHz$^{-1}$, see 
Fig.~\ref{fig:si_plots}, bottom panel) 
and the absence of emission lines in the optical spectra may suggest that these 
objects belong to the class of FRI 
radio galaxies; but FRI radio galaxies are characterised, on average, by 
steeper radio spectra and larger linear sizes (but see the linear size -- 
radio power relation found for B2 radio galaxies, \citealt{deRuiter1990} and 
references therein).

The compactness of the sources, together with the
flat/inverted spectra, suggests core emission with strong synchrotron or 
free-free self-absorption. This could be associated to either very early 
phases of nuclear radio-activity (the so-called GHz peaked spectrum - GPS - 
radio sources, \citealt{Odea1998,Snellen2000}) or late phases of the 
evolution of 
AGNs, characterised by low accretion/radiative efficiency 
(advection-dominated accretion flow, i.e. 
ADAF; advection dominated inflow-outflow solutions, i.e. ADIOS). 
In the first case, however, larger luminosities are expected 
($P_{1.4 \rm{GHz}}>10^{25}$ WHz$^{-1}$), while in the latter case 
very low radio powers are predicted ($P_{5 \rm{GHz}}<10^{21}$ WHz$^{-1}$; 
see \citealt{Doi2005}). 
Another intriguing possibility is that in these sources ADAF and radio jets 
coexist, as suggested for low luminosity AGNs, (LLAGNs, see e.g. 
\citealt{Doi2005} and reference therein). 
This would explain the somewhat brighter
luminosities than expected by simple ADAF and can still be consistent with the 
presence of flat/inverted radio spectra (see ADAF-jet model by 
\citealt{Falcke1999}). \\

This class of objects may also be very similar to the composite 
class of the so-called low power ($P_{408 \rm{MHz}}<10^{25.5}$ WHz$^{-1}$) 
compact ($<10$ kpc) -- LPC -- radio sources studied by \cite{Giroletti2005}. 
Their host galaxies do not show signatures of strong nuclear
activity in the optical (and X-ray) bands. 
Preliminary results indicate that multiple causes can produce LPC sources:
geometrical-relativistic effects (low power BL-Lacertae objects), youth, 
instabilities in the jets, frustration by a denser than average ISM,  and a
premature end of nuclear activity.  

\section{Summary}\label{sec:summary}

In this paper we have discussed the nature of the faint, sub-mJy, radio 
population,
using a sample of 131 radio sources that were observed at
1.4 and 5 GHz with the ATCA (the ATESP--DEEP1 sample). 
A smaller sample of 85 radio sources
is covered by deep multi--colour images.
These were optically identified down 
to very faint magnitudes, which was possible thanks
to the availability of very deep multi--colour optical material 
(in U, B, V, R, I, and sometimes J and K bands).
The high percentage of identifications ($\sim 78\%$) makes this a sample 
that is well suited for follow up studies
concerning the composition of the sub-mJy population and, in general, the 
cosmological evolution of
the various classes of objects associated with faint radio sources.

We summarise our main results here.
\begin{itemize}
\item For  85\% of the identification sample we succeeded in 
deriving reliable photometric 
redshifts, based on the available accurate colours (UBVRIJK).
\item Based on spectral types determined either directly from spectroscopy or 
from the photometry (or both), we find that at the sub-mJy level the large 
majority of sources are associated with objects that have
early type (64\%) and  AGN (14\%) spectra; these are of course what we would 
normally call radio galaxies and quasars. 
\item Although earlier work (based on shallower optical imaging and 
spectroscopy) revealed the presence of a conspicuous component of 
late type and
star-burst objects, such objects appear to be important only at brighter 
magnitudes ($I<19$), and are rare at fainter magnitudes ($19<I<23.5$).
\item From an overall comparison of the radio spectral index with other 
radio and optical properties of the entire ATESP--DEEP1 sample,
we find that most sources with flat radio spectra have high radio-to-optical 
ratios, as expected for classical radio
galaxies and quasars. Flat-spectrum sources with low radio-to-optical ratios 
are preferentially associated with ETS, in which the 
radio 
emission is most plausibly triggered by nuclear activity as well, while 
star-forming galaxies are associated to steep-spectrum radio sources.
\item ETS with flat or inverted spectra are mostly compact, 
with linear size $<10-30$ kpc, suggesting core-dominated radio emission. Their
low radio luminosities (in the range $10^{22}-10^{24}$ W/Hz at 
1.4 GHz) and the absence of emission lines
in their spectra (when available) suggest that they are FRI sources, 
although these would normally have 
steeper spectra and be more extended. They may therefore represent specific 
phases in the life of a radio source, or may be similar to the low power 
compact radio sources discussed by \cite{Giroletti2005}. 
\end{itemize}

\begin{acknowledgements}
AM thanks Luiz da Costa and the ESO Imaging Survy (EIS) Team
for their hospitality and for the assistance with the optical and
NIR data reduction during his stay in Garching.
\end{acknowledgements}

\tabcolsep 0.2cm
\begin{table*}
\caption{List of the optical identifications of ATESP radio sources in fields 
DEEP1a, b and c. We list in 
Col.~1 the source name and in Cols.~2--8 
the $U,B,V,R,I,J,K_{s}$--band magnitudes (Vega system), when available. In Cols.
 9 to 11 we list the photometric redshift, together with the corresponding SED and 
age (for SED templates, types and ages we refer to {\it Hyperz} 
documentation; we note that CWW and quasar templates do not provide ages).
Spectroscopy information, when available, is reported in Cols. 12 
(redshift) and 13 (spectral types). References and notes  
are listed in  Col. 14.
}
\label{tab:LRidentification}
\scriptsize
\begin{tabular}{lr|rrrrrrr|rcc|ccc|c}
\hline
\hline
\multicolumn{2}{c}{Source Name} &  U & B & V & R & I & J & K$_s$ & $z_{phot}$ 
& SED & Age (Gyr) & $z_{sp}$ & Sp. Type & Notes & Class\\
\hline
&  & & & & & & & & & & & & & & \\
ATESP5  &J224750-400148 &	$>25.1$&    21.8 	&  20.7  &  	19.3	&	18.4	&    -  &-   & 0.35 & Burst & 8.5 & 0.442 & ETS & (a) & ETS\\
ATESP5	&J224753-400455 &	$>25.1$&    23.4 	&  22.5  &  	21.3	&	20.1	&    -  & -   & 0.61 &  Burst & 1.7 &  - &  - &  & ETS \\
ATESP 	&J224759-400825 &	$>25.1$&    25.4 	&  25.2  &  	25.2	&	24.3	&    -  & - &  - & -   & - & - & - & & 	 -   	\\
ATESP5	&J224801-400542 &	$>25.1$&    23.1 	&  22.1  &  	20.9	&	19.8	&    -       &	-   & 0.56 & Burst & 2.0 & - & - &  & ETS \\
ATESP 	&J224803-400513 &	17.7	&    18.4 	&  17.9  &  	17.6	&	17.2	&    -  & -   & 1.00 & QSO & - & 1.72 & QSO & (bc) & AGN\\
ATESP5	&J224809-402211 &	22.5	&    23.0 	&  22.7  &  	21.5	&	20.8	&    -       &	-   & 0.55 & Sc & 4.5 & - & - &	& LTS   \\
ATESP 	&J224811-402455 &	22.0	&    22.7 	&  22.4  &  	21.6	&	21.1	&    -       &	-   & 0.55 & Burst & 0.05 & - & - && LTS \\
ATESP 	&J224817-400819 &	$>$25.1&    23.9 	&  22.9  &  	21.8	&	20.3	&    -       &	-   & 0.80 & Burst & 1.0 & - & - & & ETS \\
ATESP5	&J224822-401808 &	$>$25.1&    23.5 	&  22.2  &  	20.6	&	19.4	&    -       &	-   & 0.37 & Ell & 6.5 & - & - & & ETS 	 \\
ATESP 	&J224843-400456 &	24.3	&    25.6 	&  24.8  &  	24.9	&	24.3	&    -       &	-  &  - & -   & - & - & - &  & -   	\\
ATESP5	&J224850-400027 &	21.8	&    22.3	&  21.8  &  	20.8	&	19.7	&    -       &	-   &  - & -   & - & - & - & & -	 \\
ATESP5	&J224858-402708 &	22.1	&    22.3 	&  21.6  &  	20.5	&	19.8	&    -       &	-   & 0.48 & Burst & 0.09 & - & - && LTS  \\
ATESP 	&J224911-400859 &	17.8	&    17.4 	&  16.9  &  	16.2	&	15.6	&    -  & -   & 0.11 & Sbc & - & 0.065 & LTS & (b) & LTS\\
ATESP5	&J224919-400037 &	$>$25.1&    22.6 	&  21.5  &  	20.0	&	19.0	&    -       &	-   & 0.35 & Burst & 10.5 & - & - && ETS    \\
ATESP5	&J224932-395801 &	23.6	&    24.1 	&  23.0  &  	21.6	&	20.4	&    -  & -   & 0.60 & Burst & 2.0 & 0.713 & ETS & (ad) & ETS\\
ATESP5	&J224935-400816 &	18.3	&    17.6 	&  16.6  &  	15.9	&	15.2	&    -  & -  & 0.11 & Burst & 12.5 & 0.153  & ETS & (a) & ETS\\
ATESP5	&J224948-395918 &	$>$25.1&    25.3 	&  25.9  &  	24.5	&	23.5	&    -       &	-   & - & - & - & - & - & & -	     	\\
ATESP5	&J224951-402035 &	14.6	&    18.6 	&  18.4  &  	18.0	&	17.6	&    -       &	-   & - & - & - & - & - & & -	     	\\
ATESP5	&J224958-395855 &	13.6	&    16.7 	&  16.4  &  	16.3	&	15.2	&    -       &	-   & - & - & - & 0.249  & ETS & (b) & ETS\\
ATESP5	&J225004-402412 &	$>$25.1&    23.4 	&  22.3  &  	21.2	&	20.0	&    -       &	-   & 0.61  & Burst & 1.7 & - & - & & ETS\\
ATESP5	&J225008-400425 &	18.5	&    18.0 	&  17.1  &  	16.4	&	15.7	&    -  & -   & 0.10 & Burst & 3.5 & 0.126  & ETS& (e) & ETS \\
ATESP 	&J225009-400605 &	$>$25.1&    25.1 	&  23.8  &  	22.4	&	21.0	&    -       &	-   & 0.74   & Burst & 1.0 & - & - & & ETS\\
ATESP5	&J225028-400333 &	20.7	&    21.0 	&  20.4  &  	19.2	&	18.5	&    -  &-   & 0.51 & Burst & 0.18 & 0.540  & LTS & (b) & LTS\\
ATESP5	&J225048-400147 &	$>$24.6&    $>$25.7 	&  26.4  &  	25.9	&	23.8	&    -       &	-   & - & - & - & - & - & & -	     	\\
ATESP5	&J225056-400033 &	23.9	&    24.7 	&  24.6  &  	23.2	&	22.2	&    -       &	-   & 1.43 & REDQ & - & - & - & & AGN	  \\
ATESP5	&J225056-402254 &	21.2	&    20.4 	&  19.1  &  	18.3	&	17.4	&    -   & -   & 0.21 & Burst & 11.5 & - & - & & ETS\\
ATESP5	&J225057-401522 &	15.8	&    15.3 	&  14.6  &  	14.0	&	13.2	& -  & 13.0  & 0.01 & Burst & 0.72 & 0.033 & ETS & (af) & ETS\\
ATESP5	&J225058-401645 &	$>$24.6&    26.3	&  24.7  &  	23.1	&	21.0	&    -       &	18.6   & 0.96 & Burst & 1.02 & - & - & & ETS \\
ATESP5	&J225100-400934 &	$>$24.6&    $>$25.7 	&  26.9  &  	24.5	&	22.5	&    -       &	18.0   & 1.21 & Ell & 5.5 & - & - & & ETS\\
ATESP5	&J225112-402230 &	26.0	&    $>$25.7 	&  27.8  &  	26.5	&	24.4	&    -       &	18.5   & - & - & - & - & - &  & -   	\\
ATESP5	&J225122-402524 &	23.1	&    23.7 	&  23.5  &  	23.1	&	22.6	&    -       &	19.5   & 2.23 & SB2 & - & - & - & & SB	\\
ATESP5	&J225138-401747 &	19.4	&    19.3 	&  18.4  &  	17.8	&	17.0& 16.0 & 14.5 & 0.21 & Burst & 0.26 & 0.235 & LTS & (e) & LTS \\
ATESP 	&J225206-401947 &	20.3	&    20.6 	&  20.1  &  	19.7	&	19.1	&    18.4 &  17.1   & 2.06 & QSO & - & - & - & & AGN	 \\
ATESP5	&J225217-402135 &	22.7	&    23.8 	&  23.4  &  	22.9	&	22.3	&    20.9 &  19.2   & 0.93 & BALQ & - &	- & - & & AGN   \\
ATESP5	&J225223-401841 &	15.7	&    15.4 	&  14.9  &  	14.3	&	13.6	&  12.7 & 11.5 & 0.04 & Sa & 12.5 & 0.033 & ETS & (af) & ETS\\
ATESP5	&J225239-401949 &	14.9	&    14.8 	&  14.3  &  	13.8	&	13.1	&    12.2 &  11.3 & 0.06 & Sa & 8.5 &0.033 & ETS & (e)& ETS \\
ATESP5	&J225242-395949 &	23.9	&    24.2 	&  22.9  &  	21.7	&	20.7	&    -       &	-   & 0.41 & Burst  & 0.36 & - & - & & ETS \\
ATESP5	&J225249-401256 &	24.0	&    25.7 	&  24.3  &  	22.6	&	20.9	&    19.4 &  17.5   & 0.59 & Ell & 5.5 &  - & - & & ETS \\
ATESP5	&J225316-401200 &	23.7	&    23.6 	&  22.5  &  	21.2	&	20.0	&    -       &	-   & 0.36 & Ell & 5.5 & - & - & & ETS 	 \\
ATESP5	&J225321-402317 &	22.1	&    23.3 	&  22.8  &  	21.9	&	21.2	&    -       &	-   & 0.70 & Burst & 0.18 & - & - & & LTS\\
ATESP5	&J225322-401931 &	$>$24.6&    25.3 	&  23.1  &  	21.7	&	20.0	&    -       &	-   & 0.30 & Burst & 10.5 & - & - & & ETS\\
ATESP5	&J225323-400453 &	$>$24.6&    25.0	&  24.0  &  	22.8	&	21.5	&    -       &	18.2   & 0.36 & Sa & 6.5 & - & - & & ETS  \\
ATESP5	&J225325-400221 &	23.1	&    22.8 	&  21.7  &  	20.4	&	19.1	&    -       &	16.1   & 0.37 & Ell & 5.5 & - & - & & ETS    \\
ATESP5	&J225332-402721 &	$>$24.6&    25.6 	&  26.9  &  	24.2	&	24.0	&    -       &	-   & - & - & - & - & - & & -	     	\\
ATESP5	&J225344-401928 &	23.8	&    24.9 	&  24.4  &  	24.1	&	22.7	&    -       &	18.8   & 1.40 & Ell & 3.5 & - & - & & ETS \\
ATESP5	&J225345-401845 &	19.9	&    19.2 	&  18.1  &  	17.3	&	16.3	&    -       &	14.3   & 0.29 & Burst & 1.7 & - & - & & ETS\\
ATESP 	&J225351-400441 &	$>$25.3&    23.9 	&  23.7  &  	23.2	&	22.4	&    -       &	-   & 2.20 & QSO & - & - & - &	& AGN    \\
ATESP5	&J225353-400154 &	21.3	&    21.3 	&  21.4  &  	20.9	&	19.7	&    -       &	17.1   & 2.00 & REDQ & - & - & - & & AGN\\
ATESP5	&J225400-402204 &	16.8	&    16.4 	&  15.8  &  	15.1	&	14.2	&    - & - & 0.03 & Sa & 13.5 & 0.033 & LTS & (g) & LTS\\
ATESP5	&J225404-402226 &	21.4	&    21.4 	&  20.7  &  	19.5	&	18.6	&    -       &	16.2   & 0.43 & Ell & 4.5 & - & - & &ETS\\
ATESP5	&J225414-400853 &	15.1	&    15.3 	&  15.0  &  	14.6	&	14.1& 13.6 & 12.8 & 0.07 & Burst & 0.09 & 0.032 & LTS & (e) & LTS\\
ATESP5	&J225430-400334 &	$>$25.3&    23.4 	&  22.3  &  	20.9	&	19.4	&    -       &	16.4   & 0.56 & Burst & 2.6 & - & - && ETS \\
ATESP 	&J225430-402329 &	$>$25.3&    25.6 	&  25.1  &  	24.1	&	22.3	&    -       &	18.7   & 1.03 & Ell & 4.5 & - & - & & ETS\\
ATESP5	&J225434-401343 &	$>$25.3&    25.8 	&  25.8  &  	24.6	&	23.2	&    20.7 &  19.0   & 1.64 & Burst & 0.36 & - & - & & ETS\\
ATESP5	&J225436-400531 &	$>$25.3&    22.2 	&  21.1  &  	19.7	&	18.4	&    17.7 &  16.0   & 0.44 & Burst & 4.5 & - & - & & ETS\\
ATESP5	&J225442-400353 &	$>$25.3&    25.1 	&  24.1  &  	22.9	&	21.0	&    19.5 &  17.6   & 0.81 & Ell & 5.5 & - & - & & ETS	   \\
ATESP5	&J225449-400918 &	23.2	&    23.6 	&  23.3  &  	23.2	&	22.3	&    20.6 &  18.9   & 1.58 & Ell & 2.6 & - & - & & ETS	  \\
ATESP5	&J225509-402658 &	21.6	&    22.3 	&  21.8  &  	21.0	&	19.4	&    -       &	16.5   & 0.97 & Burst & 0.13 & - & - & & LTS\\
ATESP5	&J225515-401835 &	23.3	&    23.7 	&  23.0  &  	22.5	&	21.1	&    19.2 &  15.9   & 2.20 & REDQ & - & - & - &	 & AGN    \\
ATESP5	&J225529-401101 &	$>$25.3&    23.3 	&  22.2  &  	21.0	&	19.5	&    18.7 &  16.6   & 0.73 & Burst & 0.36 & - & - & & ETS  \\
&  & & & & & &   & & & & & &  &  &\\
\hline
&  & & & & & &    & & & & & &  &  &\\
ATESP5	&J224827-402515 &	22.2 &    22.8   	&  22.8  &   22.0 &	21.0 &    -       &	-   & 1.25 & Burst & 0.36 & - & - & & ETS	    \\
ATESP5	&J225034-401936 &	$>$24.6&    24.5 	&  24.0  & 	22.9	&	21.1	&    -       &	-   & 1.17 & Burst & 0.72 & - & - & & ETS\\
ATESP5	&J225426-402442 &	$>$25.3&    24.1 	&  23.9  &  	23.4	&	22.2	&    -       &	18.7   & 1.92 & REDQ & - & - & - & & AGN\\
ATESP5	&J225443-401147 &	$>$25.3&    $>$25.9   &  $>$25.8 &   $>$25.7&	$>$23.8&    $>$22.2&  20.0   & - & - & - & - & - && -		\\
ATESP5	&J225505-401301 &	20.6	&    19.8	&  18.6  &  	17.7	&	16.8	&    16.0 &  14.7   & 0.35 & Burst & 1.0 & - & - & & ETS    \\
ATESP5	&J225511-401513 &	$>$25.3&    $>$25.9   &  $>$25.8 &   $>$25.7&	$>$23.8&    21.2 &  18.3   &- &- &- & - & - &&-		\\
&  & & & & & &   & & & &  & & & & \\
\hline
\multicolumn{16}{l}{$^{a}$ On-going spectroscopy program (Prandoni et al. in 
prep.)}\\
\multicolumn{16}{l}{$^{b}$ Spectroscopy from \cite{Prandoni2001b}.}\\
\multicolumn{16}{l}{$^{c}$ Spectroscopic redshift re-measured: published 
value was over-estimated.}\\
\multicolumn{16}{l}{$^{d}$ Same redshift obtained by \cite{Olsen2005}: 
galaxy in a cluster at $z=0.710$.}\\
\multicolumn{16}{l}{$^{e}$ Spectroscopy from \cite{Vettolani98}.}\\
\multicolumn{16}{l}{$^{f}$ Same redshift obtained by \cite{DiNella1996}.}\\
\multicolumn{16}{l}{$^{g}$ Spectroscopy obtained from 6dF public data.}\\
\end{tabular}
\end{table*}

\bibliographystyle{aa}
\bibliography{atesp_v4}

\begin{thebibliography}{}
\addcontentsline{toc}{chapter}{Bibliography}

\bibitem[Afonso \etal (2006)]{Afonso2006}
Afonso J., Mobasher B., Koekemoer A., Norris R.P., Cram L., 2006, AJ, 131, 1216

\bibitem[Bertin \& Arnouts (1996)]{Bertin1996}
Bertin E., Arnouts S., 1996, A\&AS, 117, 393

\bibitem[Bolzonella \etal (2000)]{Bolzonella2000}
Bolzonella M., Miralles J.-M., Pell\'o R., 2000, A\&A, 363, 476

\bibitem[Bondi \etal (2003)]{Bondi2003}
Bondi M., Ciliegi P., Zamorani G., et al., 2003, A\&A, 403, 857

\bibitem[Bruzual \& Charlot (1993)]{BruzualCharlot1993}
Bruzual G., Charlot S., 1993, ApJ, 405, 538

\bibitem[Ciliegi \etal (2005)]{Ciliegi2005}
Ciliegi P., Zamorani G., Bondi M., et al., 2005, \aap, 441, 879

\bibitem[Ciliegi \etal (2003)]{Ciliegi2003}
Ciliegi P., Zamorani G., Hasinger G., et al., 2003, A\&A, 398, 901

\bibitem[Coleman \etal (1980)]{Coleman1980}
Coleman D.G., Wu C.C., Weedman D.W., 1980, ApJS, 43, 393

\bibitem[Condon (1980)]{Condon1980}
Condon J.J., 1980, ApJ, 242, 894

\bibitem[Condon \& Broderick (1988)]{Condon1988}
Condon J.J., \& Broderick J.J., 1988, AJ, 96, 30

\bibitem[de Ruiter \etal (1977)]{deRuiter1977}
de Ruiter H.R., Willis A.G., Arp H.C., 1977, A\&AS, 28, 211 
1997, A\&A, 319, 7

\bibitem[de Ruiter \etal (1990)]{deRuiter1990}
de Ruiter H.R., Parma P., Fanti C., Fanti R., 
1990, A\&A, 351, 361

\bibitem[Di Nella \etal (1996)]{DiNella1996}
Di Nella H., Couch W.J., Paturel G., and Parker Q.A., 1996, MNRAS, 283, 367

\bibitem[Doi \etal (2005)]{Doi2005}
Doi A., Kameno S., Kohno K., Nakanishi K., Inoue M., 2005, MNRAS, 363, 692

\bibitem[Donnelly \etal (1987)]{Donnelly1987}
Donnelly R.H., Partridge R.B., Windhorst R.A., 
1987, ApJ, 321, 94

\bibitem[Erben \etal (2005)]{Erben2005}
Erben T., Schirmer M., Dietrich J.P., et al., 2005, Astronomische
Nachrichten, 326, 432

\bibitem[Falcke \& Bierman (1999)]{Falcke1999}
Falcke \& Biermann 1999, A\&A, 342, 49

\bibitem[Fanaroff \& Riley (1974)]{Fanaroff1974}  
Fanaroff B.L., Riley J.M., 1974, MNRAS, 167P, 31F

\bibitem[Georgakakis \etal (1999)]{Georgakakis1999}
Georgakakis A., Mobasher B., Cram L., et al., 1999, MNRAS 306, 708

\bibitem[Girardi \etal (2005)]{Girardi2005}
Girardi L., Groenewegen M., Hatziminaoglou E., da~Costa L., 2005, A\&A,
  436, 895

\bibitem[Giroletti \etal (2005)]{Giroletti2005}
Giroletti M., Giovannini G., Taylor G.B., 2005, A\&A, 441, 89

\bibitem[Gruppioni \etal (1999)]{Gruppioni1999}
Gruppioni C., Ciliegi P., Rowan-Robinson M., et al., 1999, MNRAS 305, 297

\bibitem[Gruppioni \etal (1997)]{Gruppioni1997}
Gruppioni C., Zamorani G., de Ruiter H.R., et al., 1997, MNRAS 286, 470

\bibitem[Hatziminaoglou \etal (2000)]{Hatziminaoglou2000} 
Hatziminaoglou E., Mathez G., Pell\'o R., 2000,
A\&A, 359, 9

\bibitem[Hildebrandt \etal (2006)]{Hildebrandt2006} 
Hildebrandt H., Erben T., Dietrich J.P., et al. 2006,
A\&A, 452, 1121

\bibitem[Hopkins \etal (1998)]{Hopkins1998}
Hopkins A.M., Mobasher B., Cram L., Rowan-Robinson M.,
1998, MNRAS, 296, 839

\bibitem[Magliocchetti \etal (2000)]{Magliocchetti2000}
Magliocchetti M., Maddox S.J., Wall J.V., et al., 2000, MNRAS, 318, 1047

\bibitem[Mignano \etal (2007)]{Mignano2007}
Mignano A., Miralles J.-M., da Costa L., et al., 2007, A\&A, 462, 553

\bibitem[Nonino \etal (1999)]{Nonino99}
Nonino M., Bertin E., da Costa L., et al., 1999, A\&AS 137, 51 

\bibitem[O'Dea (1998)]{Odea1998}
O'Dea 1998, PASP, 110, 493

\bibitem[Olsen \etal (2005)]{Olsen2005}
Olsen L.F., Zucca E., Bardelli S., et al., 2005, A\&A, 442, 841

\bibitem[Olsen \etal (2006)]{Olsen2006}
Olsen L.F., Miralles J.-M., da Costa L., Vandame B., Madejsky R., 
Jorgensen H.E., Mignano A., et al., 
2006, A\&A, 456, 881

\bibitem[Prandoni \etal (2000a)]{Prandoni2000a}
Prandoni I., Gregorini L., Parma P., 
de Ruiter H.R., Vettolani G., Wieringa M.H., Ekers R.D., 
2000a, A\&AS, 146, 31

\bibitem[Prandoni \etal (2000b)]{Prandoni2000b}
Prandoni I., Gregorini L., Parma P., 
de Ruiter H.R., Vettolani G., Wieringa M.H., Ekers R.D., 
2000b, A\&AS, 146, 41 

\bibitem[Prandoni \etal (2001)]{Prandoni2001b}
Prandoni I., Gregorini L., Parma P.,  
de Ruiter H.R., Vettolani G., Zanichelli A., Wieringa M.H., Ekers R.D., 
2001, A\&A, 369, 787

\bibitem[Prandoni \etal (2006)]{Prandoni2006}
Prandoni I., Parma P., Wieringa M.H., de Ruiter H.R., Gregorini L., 
Mignano A., Vettolani G., Ekers R.D., 
2006, A\&A, 457, 517, Paper I

\bibitem[Reichard \etal (2003)]{Reichard2003b}
Reichard T.A., Richards G.T., Schneider D.P., Hall P.B., Tolea A., 
Krolik J.H., Tsvetanov Z., Vanden Berk D.E., et al., 
2003, \aj, 125, 1711

\bibitem[Richards \etal (1999)]{Richards1999}
Richards E.A., Fomalont E.B., Kellermann K.I., et al., 1999, ApJ 526, L73

\bibitem[Richards \etal (2003)]{Richards2003}
Richards G.T., Hall P.B., Vanden Berk D.E., Strauss M.A., Schneider D.P., 
Weinstein M.A., Reichard T.A., York D.G., Knapp G.R., Fan Xiaohui, 
2003, AJ, 126, 1131

\bibitem[Rigby \etal (2007)]{Rigby2007}
Rigby E.E., Snellen I.A.G., Best P.N., 2007, MNRAS, in press 
(arXiv:0706.2323 [astro-ph])

\bibitem[Schinnerer \etal (2006)]{Schinnerer2006}
Schinnerer E., Smolcic V., Carilli C.L., {et~al.} 2006, ApJS, COSMOS special 
issue

\bibitem[Schirmer \etal (2003)]{Schirmer2003}
Schirmer M., Erben T., Schneider P., {et~al.} 2003, \aap, 407, 869

\bibitem[Simpson \etal (2006)]{Simpson2006}
Simpson C., Martinez-Sansigre A., Rawlings S. et al., 2006, MNRAS, 372, 741

\bibitem[Snellen \etal (2000)]{Snellen2000}  
Snellen I.A.G., et al., 2000, MNRAS 319, 445

\bibitem[Snellen \etal (1996)]{Snellen1996}  
Snellen I.A.G., Bremer M.N., Schilizzi R.T., Miley G.K., van Ojik R., 1996, 
MNRAS 279, 1294

\bibitem[Sullivan \etal (2004)]{Sullivan2004}
Sullivan M., Hopkins A.M., Afonso J., Georgakakis A.
Chan B., Cram L.E., Mobasher B., Almeida C.,
2004, ApJS, 155, 1

\bibitem[Sutherland \& Saunders (1992)]{Sutherland1992}
Sutherland W. \& Saunders W., 1992, MNRAS, 259, 413

\bibitem[Vettolani \etal (1997)]{Vettolani97}
Vettolani G., Zucca E., Zamorani G., et al., 1997, A\&A 325, 954

\bibitem[Vettolani \etal (1998)]{Vettolani98}
Vettolani G., Zucca E., Merighi R., et al., 1998, A\&AS 130, 323

\bibitem[Willott \etal (2003)]{Willott2003}
Willott C.J., Rawlings S., Jarvis M.J., Blundell K.M., 2003,
MNRAS, 339, 173

\bibitem[Windhorst \etal (1990)]{Windhorst90}
Windhorst R.A., Mathis D., Neuschaefer L., 1990. In: Kron R.G. (ed.) 
Evolution of the Universe of Galaxies, ASP Conf. Ser. 10, 389

\bibitem[York \etal (2000)]{York2000}
York, D. G., Adelman, J., Anderson, J. E. et~al., 2000 AJ, 120, 1579


\end{thebibliography}




















\end{document}